\documentclass[aps,pre, preprint,groupedaddress,amsmath,amssymb,showpacs,eqsecnum]{revtex4}

\usepackage{bm}
\usepackage{graphicx}
\usepackage{subfigure}

\newcommand{\Dx}{\int \mathcal{D}x \, \rho_y[x]\,}
\newcommand{\Dc}{\int \mathcal{D}c \, \rho_{(Q,\mathbf{q})}[c]\,}
\newcommand{\vecr}{\mathbf{r}}
\newcommand{\vrr}{\mathbf{r'}}
\newcommand{\vrn}[1]{\mathbf{r}_{#1}}
\newcommand{\vrrn}[1]{\mathbf{r}_{#1}'}
\newcommand{\ir}{\int_V d^3r \,}
\newcommand{\irr}{\int_V d^3r' \,}
\newcommand{\irn}[1]{\int_V d^3r_{#1} \,}
\newcommand{\irrn}[1]{\int_V d^3r_{#1}' \,}
\newcommand{\pr}{\Pi_{\rho}}
\newcommand{\pg}{\bm{\Pi}_{\mathbf{g}}}
\newcommand{\pe}{\Pi_{\epsilon}}

\newcommand{\pQ}{\Pi_Q}
\newcommand{\pq}{\bm{\Pi}_{\mathbf{q}}}
\newcommand{\pil}{\Pi_l}
\newcommand{\ke}{\kappa_{\rm E}}
\newcommand{\kb}{k_{\rm B}}

\usepackage{ulem}
\newcommand{\asja}[1]{{#1}}

\begin{document}

\title
{Bridging length and time scales in sheared demixing systems: from the Cahn-Hilliard to the Doi-Ohta model}

\author{Asja Jeli\'c}
\altaffiliation[Present address: ]
{Universit\'e Pierre et Marie Curie - Paris VI, LPTHE UMR 7589, 
4 Place Jussieu, 75252 Paris Cedex 05, France}, 
\email{asja@lpthe.jussieu.fr}
\author{Patrick Ilg} 
\author{Hans Christian \"Ottinger}
\affiliation{Polymer Physics, Department of Materials, ETH Zurich, 8093 Zurich, Switzerland}

\date{\today}

\begin{abstract}
We develop a systematic coarse-graining procedure which establishes 
the connection between models 
of mixtures of immiscible 
fluids at different length and time scales. 
We start from the Cahn-Hilliard model of spinodal decomposition in a
binary fluid mixture under flow from which we derive the coarse-grained 
description. 
The crucial step in this procedure is to identify the relevant 
coarse-grained variables and find the appropriate mapping which 
expresses them in terms of the more microscopic variables.
In order to capture the physics of the Doi-Ohta level, we introduce the 
interfacial width as an additional variable at that level. In this way, we
account for the stretching of the interface under flow
and derive analytically the convective behavior of the relevant
coarse-grained variables, which in the long wavelength limit
recovers the familiar phenomenological Doi-Ohta model.
In addition, we obtain the expression for the interfacial tension in terms of
the Cahn-Hilliard parameters as a direct result of the
developed coarse-graining procedure. Finally, by
analyzing the numerical results 
obtained from the simulations on the Cahn-Hilliard level, 
we discuss that  
dissipative processes at the Doi-Ohta level are of the
same origin as in the Cahn-Hilliard model.
The way to estimate the
interface relaxation times of the Doi-Ohta model from the
underlying morphology dynamics simulated at the Cahn-Hilliard level
is established.
\end{abstract}


\maketitle

\section{Introduction}\label{sec:Introduction}
The problem of coarse graining in terms of 
bridging time and length scales between microscopic and macroscopic
levels of description is a crucial issue in the physics of complex
fluids, like polymer melts, colloids, liquid crystals and emulsions,
out of equilibrium. The wide span of length and time scales is particularly 
evident in these systems, due to their internal structure leading to additional 
mesoscopic levels of description, intermediate between microscopic and
macroscopic \cite{Larson_99}. Many of the
practical applications of these fluids crucially depend on the
evolution of their complex, multiphase
morphologies developed through equilibrium self-assembling or during
non-equilibrium processing of these systems.
Some examples of such applications are food processing, membrane 
technology, encapsulation systems, drug delivery, coating, production 
of paint and cosmetics etc \cite{Larson_99,complex_fluids-1,complex_fluids-2,complex_fluids-3}. 
Therefore, 
the goal of connecting different levels 
of description of complex fluids is of great importance. In the present work,
this problem is approached by considering the phase separation of binary 
fluid mixtures subjected to a shear flow.

When a binary (AB) fluid mixture is quenched from a high temperature
homogeneous phase to an unstable region below the coexistence curve, it 
becomes unstable with respect to long-wavelength fluctuations in the composition 
and starts to phase separate \cite{Bray-94}.  
The interface between the A- and B-rich domains is initially very diffuse, but sharpens 
with time. In the later stages of 
phase separation, local equilibrium is achieved within each domain and the effects
of interfacial energy become important. The domain pattern further coarsens 
in time, while the dynamics is governed by the minimization of the excess 
interfacial energy of the system. The kinetics of this nonequilibrium phenomena
is an area of extensive research \cite{Bray-94}. Especially the effects of a shear 
flow on the deformation and kinetics of domain growth, and the corresponding
rheological behavior, are challenging and technologically important
topics \cite{Onuki-97R}. The complexity of the problem lies in the
presence of a complex interface and its motions due to coagulation,
rupture and deformation of domains, which significantly influences
the macroscopic properties of a mixture. Although numerous experimental 
\cite{KrSengHam_92,Lauger-95,Matsuzaka-98,Derks-08,Aarts-05}, numerical \cite{OhtaNozDoi_90,PadTox_97,CorGonLam-1,CorGonLam-2,CorGonLam-3,CorGonLam-4,Zhang_00,Berth_01, Fielding_08,ShouChakr_00,NESS-1,NESS-2,NESS-3} and 
analytical \cite{Onuki_87,DoiOhta_91,CorGonLam-1,CorGonLam-2,CorGonLam-3,CorGonLam-4,Fielding_08,CH-analytical-1,CH-analytical-2} studies have 
been done in order to understand the flow effects 
on the morphology and the dynamics in this system, many questions still 
remain open, like the possibility of achieving a nonequilibrium steady 
state \cite{ShouChakr_00,NESS-1,NESS-2,NESS-3, Fielding_08}.

Due to the complex
morphology of this system, even in the case of two simple Newtonian
fluids, one can identify several length and time scales requiring 
different theoretical approaches and simulation methods. The 
standard mesoscopic model for understanding the dynamics of spinodal
decomposition is based on the formulation of Cahn and Hilliard
\cite{CaHil_58}. It describes the kinetics of the system in terms of
the convective-diffusion equation for the order parameter - composition
$c$, and the Navier-Stokes equation for the fluid velocity $\mathbf{v}$ 
(so-called `model H', see e.g.\ \cite{model-H}):
\begin{subequations}\label{eq:Cahn-Hilliard}
\begin{eqnarray}
   \frac{\partial c}{\partial t}+\mathbf{v}\cdot\left( \nabla c \right)
    &=&M \nabla^2 \mu_c, \label{eq:Cahn-Hilliard-c}\\
   \rho\left( \frac{\partial \mathbf{v}}{\partial t}+\left(\mathbf{v}\cdot \nabla  \right) \mathbf{v} \right)
   &=&\eta\nabla^2\mathbf{v}-\nabla p + \mu_c \nabla c \label{eq:Cahn-Hilliard-v},
\end{eqnarray}
\end{subequations}
where $M$ is the mobility coefficient, $p$ the pressure, $\rho$ the mass
density, and $\eta$ the viscosity. The chemical potential $\mu_{c}$ is given as 
$\mu_c=\delta F \left[c\right]/\delta c$, with the free energy functional of the 
Ginzburg-Landau form $F\left[c\right]=\int d^3r\left( -(a/2)c^2+
(b/4)c^4+(\kappa /2)\left| \nabla c\right| ^2\right)$, $(a, b >0)$, where
the square brackets emphasize the occurence of functional integrations. 
During the late stages of phase separation, the time
evolution can be described by the change of the size and shapes of
the domains. Due to the dynamical scaling hypothesis, the system is 
described by a characteristic length scale $L(t)$, which is related to 
the average domain size. By using dimensional arguments, one can 
estimate the size of the contributions from the specific terms in the equations
\eqref{eq:Cahn-Hilliard}, which leads to the 
 distinction of the three growth regimes \cite{Bray-94,Sigg_79, Furuk_85}: 
 diffusive $L(t)\sim \left( M\sigma t\right)^{1/3}$,
viscous hydrodynamics $L(t)\sim \sigma t/\eta$, and inertial hydrodynamics 
$L(t)\sim \left(\sigma t^2/\rho\right)^{1/3}$, where $\sigma$ is the interfacial
tension. Under shear flow the situation becomes more complex since the 
domains elongate along the flow direction and 
the system becomes anisotropic and can not be described by a single length 
scale (see \cite{ShouChakr_00,CorGonLam-1,CorGonLam-2,CorGonLam-3,CorGonLam-4,Berth_01} 
and references therein).

On the other hand, for understanding the rheological properties of
multiphase systems, like polymer blends, under shear flow, Doi and Ohta 
suggested that the average information about the interface between the
two phases is sufficient. Hence, their phenomenological model
\cite{DoiOhta_91} focus on the interface, which is considered as a
zero width surface embedded in the fluid. 
The presence of the interface is then represented through two
state variables that describe the interfacial area per unit volume,
$Q$, and its anisotropy, $\mathbf{q}$, in a given flow field. They are 
defined in terms of the interface orientational distribution function 
$f\left(\mathbf{n}\right)$, where $\mathbf{n}$ is a unit vector normal 
to the interface. This is a probability density of finding the amount of 
interface with the orientation $\mathbf{n}$, so that the integral over all 
orientations gives the total amount of interface per unit volume. 
Therefore, for the state variables $Q$ and $\mathbf{q}$ one can write 
the following definitions
\begin{subequations}\label{eq:Qq_definition}
\begin{eqnarray}
    Q&=&\int f\left(\mathbf{n}\right) d^2n , \label{eq:Q_definition}\\
    \mathbf{q}&=&\int \left(\mathbf{nn}-\frac{1}{3}\mathbf{1}\right)f\left(\mathbf{n}
    \right)d^2n = \overline{\mathbf{nn}}-\frac{1}{3}Q\mathbf{1} \label{eq:q_definition}, 
\end{eqnarray}
\end{subequations}
where $\mathbf{1}$ is the unit tensor, and $\int d^2n$ denotes an
integration over the unit sphere. Introduction of the distribution function 
and further averaging over all possible orientations imply the loss of 
information of the detailed morphology of the system. This coarse-grained 
model keeps only the information about the average amount of interfacial 
area per unit volume and its orientation. However, the detailed information 
about the morphology, i.e.,\ the explicit interfacial position and orientation, 
is lost.

The time evolution of the new configurational variables $Q$ and
$\mathbf{q}$ is determined by two factors: the external flow field
which orients and enlarges the interface, and the interfacial
tension which has the opposite effect and governs the relaxation of
the interface. Therefore, the time evolution can be separated into
two parts:
\begin{subequations}\label{eq:Qq_DO_evolution}
\begin{eqnarray}
   \frac{\partial Q}{\partial t} &=& 
   \left.\frac{\partial Q}{\partial
   t}\right|_{\rm convection}+\left.\frac{\partial Q}{\partial
   t}\right|_{\rm relaxation}, \label{eq:Q_DO_evolution}\\
   \frac{\partial \mathbf{q}}{\partial t} &=& 
   \left.\frac{\partial \mathbf{q}}{\partial
   t}\right|_{\rm convection}+\left.\frac{\partial \mathbf{q}}{\partial
   t}\right|_{\rm relaxation} \label{eq:q_DO_evolution}.
\end{eqnarray}
\end{subequations}
The convective part of the time evolution can be found by considering
the convection behavior of the unit vector $\mathbf{n}$ normal to
the interface under affine deformations and volume preservation
assumptions. Then, the convection of the state variables
is derived using the above definitions \eqref{eq:Qq_definition}
\begin{subequations}\label{eq:Qq_DO_convection}
\begin{eqnarray}
   \left.\frac{\partial Q}{\partial t}\right|_{\rm convection} &=&
   -\nabla \cdot \left(Q \mathbf{v}\right) -\left(\nabla\mathbf{v}\right)^{\rm
   T}:\mathbf{q}+\frac{2}{3}Q\mathbf{1}:\left(\nabla\mathbf{v}\right)^{\rm T} ,
   \label{eq:Q_DO_convection}
 \\
   \left.\frac{\partial \mathbf{q}}{\partial t}\right|_{\rm convection} &=& 
   -\left(\nabla\mathbf{v}\right)\cdot \mathbf{q}-\mathbf{q}\cdot \left(\nabla\mathbf{v}\right)^{\rm
   T}-\frac{1}{3}Q\dot{\bm{\gamma}}-\mathbf{v} \cdot\nabla \mathbf{q}   
    +\frac{1}{3}\mathbf{1}\left(\nabla\mathbf{v}\right) :
   \mathbf{q} 
 \nonumber \\
  && +\left[\mathbf{q}+\frac{1}{9}Q\mathbf{1}\right]{\rm Tr}\left(\nabla\mathbf{v} \right)
    +\overline{\mathbf{nnnn}}:\left(\nabla\mathbf{v}\right)^{\rm T} 
    \label{eq:q_DO_convection}.
\end{eqnarray}
\end{subequations}
where $ \dot{\bm \gamma}=\nabla \mathbf{v}+\left( \nabla \mathbf{v}
\right) ^T$ is the symmetrized velocity gradient tensor \cite{WagOttEdw_99}. In the equation for the 
interfacial shape \eqref{eq:q_DO_convection}, the fourth moment of the interface normal
$\mathbf{n}$ appears, which must be expressed in terms of the state variables 
through an appropriate closure approximation in order to obtain a self-contained
set of time evolution equations. Doi and Ohta postulated the following closure
approximation
\begin{equation}\label{eq:closure_DO}
  \overline{\mathbf{nnnn}}:\left(\nabla\mathbf{v}\right)^{\rm T} =
  \frac{1}{Q}\overline{\mathbf{nn}}\;\overline{\mathbf{nn}}:\left(\nabla\mathbf{v}\right)^{\rm T}
\end{equation}
and verified its accuracy. \asja{Using definition \eqref{eq:q_definition}, the convective 
time evolution of the anisotropy $\mathbf{q}$ takes the form}
\begin{eqnarray}\label{eq:q_DO_convection_closure}
   \left.\frac{\partial \mathbf{q}}{\partial t}\right|_{\rm convection} &=& 
   -\left(\nabla\mathbf{v}\right)\cdot \mathbf{q}-\mathbf{q}\cdot \left(\nabla\mathbf{v}\right)^{\rm
   T}-\frac{1}{3}Q\dot{\bm{\gamma}}-\mathbf{v} \cdot\nabla \mathbf{q}   
    +\frac{1}{3}\mathbf{1}\left(\nabla\mathbf{v}\right) :
   \mathbf{q} 
 \nonumber \\
  && +\left[\mathbf{q}+\frac{1}{9}Q\mathbf{1}\right]{\rm Tr}\left(\nabla\mathbf{v} \right)
    +\frac{1}{Q}\left(\mathbf{q}+\frac{1}{3}Q\mathbf{1}\right)\left(\mathbf{q}+\frac{1}{3}Q\mathbf{1}\right):\left(\nabla\mathbf{v}\right). 
\end{eqnarray}
As for the relaxation in the Doi-Ohta model, 
the relaxation of the state variables can be written in a generalized form
\begin{subequations}\label{eq:Qq_DO_relaxation}
\begin{eqnarray}
   \left.\frac{\partial Q}{\partial t}\right|_{\rm relaxation} &=& 
   -\,\frac{1}{\tau_{\rm do, 1}}Q, 
   \label{eq:Q_relax_tau}
 \\
   \left.\frac{\partial \mathbf{q}}{\partial t}\right|_{\rm relaxation} &=& 
   -\,\frac{1}{\tau_{\rm do, 2}} \mathbf{q}.
\end{eqnarray}
\end{subequations}
In order to specify the empirical interface relaxation times $\tau_{\rm do, 1}$ 
and $\tau_{\rm do, 2}$, assumptions on the appropriate interface relaxation
mechanisms must be made. Finally, the full set of time evolution equations for the 
configurational variables $Q$ and $\mathbf q$, together with the equations
for the hydrodynamic variables $\rho$, $\mathbf g$ and $\epsilon$ has been
formulated within the GENERIC formalism \cite{WagOttEdw_99}, which for the 
stress tensor gives
\begin{equation}
    \bm{\Sigma}= -\bm{\tau}-\sigma
    \left(\mathbf{q}-\frac{2}{3}Q\mathbf{1}\right)-p\mathbf{1},
\end{equation}
where $\bm{\tau}$ is the viscous stress tensor. 
 
Although very crude, the ability of the Doi-Ohta model to connect the
morphology of the system to its macroscopic behavior, through the above given
set of rheological constitutive equations, is very appealing for
different theoretical and technological purposes. The model and its variants 
have been used to analyze multiphase flow of polymers, as well as simple 
fluids undergoing spinodal decomposition 
\cite{LeePark_94,DO-different-1,DO-different-2,DO-different-3,DO-different-4,
DO-different-5,DO-different-6,WagOttEdw_99,KrSengHam_92}. 
However, the need for explicit assumptions of the interfacial relaxation mechanisms
and the appropriate relaxation times, which depend in an unknown
manner on the underlying morphology, makes it difficult for wide and
easy use. Therefore, in this paper we concentrate on relating
the macroscopic Doi-Ohta model to the underlying morphology,
described by the more detailed Cahn-Hilliard model. The goal is to
establish a systematic, thermodynamically guided method which
determines the coarse-grained Doi-Ohta model of a phase separating
binary fluid from the more detailed Cahn-Hilliard model, in the
region where the considered system can be described with both
approaches. To connect these two
different levels of description, the General Equation for the
Non-Equilibrium Reversible-Irreversible Coupling (GENERIC) framework
is used \cite{hco,OttGrm-97a,OttGrm-97b}. This approach allows to specify 
the mathematical structure
which should be left invariant in the coarse-graining procedure and
hence assures thermodynamically consistent results. 

In the following, we first describe the GENERIC structure and formulate
the Cahn-Hilliard model within this framework. The coarse-graining procedure
which derives the more macroscopic Doi-Ohta model is then presented. 
In particular, we present some mathematical challenges which occur within
the coarse-graining procedure, as well as the efficient way to extract the interface 
relaxation times of the Doi-Ohta level from the underlying morphology.

\section{GENERIC formalism}

The main points of the GENERIC framework of
nonequilibrium thermodynamics \cite{hco,OttGrm-97a,OttGrm-97b}
can be summarized in the following way. In analogy to equilibrium
thermodynamics, one needs to choose a complete set of variables
$\mathbf{x}$, which describes the situation of interest to the
desired detail. For a system out of equilibrium the time evolution
of all the relevant state variables can be divided into the {\it
reversible} and the {\it irreversible} contributions
\begin{equation}\label{eq:rev-irrev}
 \dot{\mathbf x} = \dot{\mathbf x}|_{\rm rev} +\dot{\mathbf x}|_{\rm irrev} \; .
\end{equation}
These two contributions are obtained by considering separately the
energy $E$ and the entropy $S$ --- the two generators of reversible
and irreversible dynamics, respectively. This can be explained by
using the analogy with the classical Hamiltonian mechanics as
following. Using the classical Hamiltonian mechanics, the reversible
contribution of the time evolution, $\dot{\mathbf x}|_{\rm rev}$, is
related to the energy gradient by way of a Poisson operator
${\mathbf L}$, so that it is $\dot {\mathbf x}|_{\rm rev} = {\mathbf
L} \cdot (\delta E/\delta {\mathbf x})$. The Poisson bracket
associated to ${\mathbf L}$, given by $\{A,B\} = ( \delta A/\delta
{\mathbf x}, {\mathbf L} \cdot \delta B/\delta{\mathbf x})$ with
appropriate scalar product $( .,. )$, must be antisymmetric and
satisfy Lebniz' rule and the Jacobi identity, which are both
features that capture the nature of reversibility. The motivation
for formulating the irreversible part of the dynamics comes from the
reversible part. For the reversible dynamics, the energy plays a
distinct role --- it is a conserved quantity for closed systems and
it drives the reversible dynamics. Therefore, for the irreversible
dynamics the entropy presents an important quantity --- it does not
decrease for closed systems. For the GENERIC formulation, it is then
assumed that the irreversible contribution to the time evolution,
$\dot {\mathbf x}|_{\rm irrev}$, is driven by the entropy gradient,
i.e.,\ that it is of the form $\dot {\mathbf x}|_{\rm irrev} =
{\mathbf M} \cdot (\delta S/\delta {\mathbf x})$, with ${\mathbf M}$
a generalized friction matrix. This friction matrix contains
transport coefficients and relaxation times associated to the
corresponding dissipative effects. It is required for ${\mathbf M}$
to be (Onsager-Casimir) symmetric and the condition that it is
positive semi-definite ensures that $\dot S \ge 0$ is fulfilled.

From the above illustration, one concludes that the time evolution
of $\mathbf{x}$ can be expressed in terms of four ``building blocks"
$E$, $S$, $\mathbf L$ and $\mathbf M$ as
\begin{equation}\label{eq:generic}
 \dot {\mathbf x} ={\mathbf L}(\mathbf{x})\cdot \frac{\delta E(\mathbf{x})}
 {\delta \mathbf{x}}+{\mathbf M}(\mathbf{x})\cdot
 \frac{\delta S(\mathbf{x})}{\delta \mathbf{x}} \; .
\end{equation}
The two different contributions to the time evolution, reversible
and irreversible, are not independent. They are related by the two
complementary degeneracy requirements
\begin{equation}\label{eq:degeneracy}
 {\mathbf L}(\mathbf{x})\cdot \frac{\delta S(\mathbf{x})}{\delta
 \mathbf{x}}=\mathbf{0} \; , \; \; \; {\mathbf M}(\mathbf{x})\cdot
 \frac{\delta E(\mathbf{x})}{\delta \mathbf{x}}=\mathbf{0} \; .
\end{equation}
The first requirement expresses the reversible nature of the
$\mathbf L$ contribution to the dynamics, demonstrating the fact
that the reversible dynamics captured in $\mathbf L$ does not affect
the entropy functional. The second one expresses the conservation of
the total energy of an isolated system by the irreversible
contribution to the system dynamics captured in $\mathbf M$.

The presented GENERIC form of the time evolution equations for the
state variables $\mathbf{x}$ represents a mathematical structure
which guarantees that the chosen model is thermodynamically
consistent. Moreover, by concentrating on the thermodynamics building
blocks $E$, $S$, $\mathbf L$ and $\mathbf M$ rather than on the time 
evolution equations in systematic coarse-graining procedures, the obtained
coarse-grained equations are guaranteed to be also thermodynamically
admissible. 

\subsection{GENERIC formulation of the Cahn-Hilliard model}\label{subsec:CH_in_GENERIC}

In order to develop a systematic coarse-graining procedure which relates
the more detailed Cahn-Hilliard level to the more macroscopic Doi-Ohta level,
we must formulate the first one in the GENERIC structure. As a first step, we 
need to identify the list of independent variables $\mathbf{x}$ which fully describe the 
considered thermodynamic system. In addition to the hydrodynamic fields, 
the mass density $\rho$, the momentum density $\mathbf{g}$, and the internal
energy density $\epsilon$, we use the composition $c$  -- the mass fraction of 
one component -- in order to describe the configuration of the system. 

Taking into account the Cahn-Hilliard free
energy of a phase separating binary system,
we assume the total energy and entropy of the system, in terms of
the variables ${\mathbf x}=\left( \rho,{\bf g},\epsilon , c \right)$, to be
\begin{subequations}\label{eq:CH_functionals}
\begin{eqnarray}
  E^{\rm (1)}({\mathbf x}) &=&
  \ir \left( \frac{1}{2} \frac{{\mathbf g}^2}{\rho}+\epsilon+\frac{1}{2} \ke \left| \nabla c  \right|^2\right),  
  \label{eq:CH_energy_functional}
 \\
  S^{\rm (1)}({\mathbf x})&=& 
  \ir  s\left(\rho,\epsilon ,c \right) ,
  \label{eq:CH_entropy_functional}
\end{eqnarray}
\end{subequations}
where $V$ is the total volume of the system.
For simplicity, the gradient-squared term, which describes the contribution 
due to the presence of the interface, is included only in the energy functional
\eqref{eq:CH_energy_functional}, although it can be separated between both, 
energy and entropy \cite{Onu_05,Jelic_09}.
\asja{The actual functional form of the entropy density 
$s\left(\rho,\epsilon ,c\right)$
is not needed for the derivation of the Cahn-Hilliard time evolution equations in the 
GENERIC form. However, we use the assumption of local equilibrium, i.e.,\
$s\left(\rho,\epsilon ,c \right)$ has the same form as an equilibrium system at the corresponding state point. In the numerical simulations reported in this
paper, we have used the popular $c^4$-structure for the uniform part of the entropy density,
i.e.,\ $s\left(\rho,\epsilon ,c \right)=s_{0}\left(\rho,\epsilon \right)-ac^2/2+bc^4/4$, 
which presents the entropic contribution to the bulk free energy density} 
\cite{CorGonLam-1,CorGonLam-2,Berth_01,ShouChakr_00,NESS-2,NESS-3,Fielding_08}.
\asja{Other possible functional forms include the one for the van der Waals systems, 
for example Eq. $4$ in \cite{Onu_05}.}
The functional derivatives of the above energy and entropy functionals 
with respect to the independent variables $\mathbf x$ are given with
\begin{subequations}\label{eq:CH_funct_derivative}
\begin{eqnarray}
  \frac{\delta E^{\rm (1)}}{\delta {\mathbf x}} &=&
  \left(-\frac{1}{2}{\mathbf v}^2, {\mathbf v}, 1 \,, 
  -\ke \nabla^2c \right)^{\rm T}, \label{eq:CH_funct_derivative_E}\\
  \frac{\delta S^{\rm (1)}}{\delta {\mathbf x}} &=& 
  \left( \frac{\partial s\left(\rho,\epsilon, c\right)}{\partial \rho}, 
  {\mathbf 0}\, , \frac{1}{T} \,, 
  \frac{\partial s\left(\rho,\epsilon, c\right)}{\partial c}\right)^{\rm T} \label{eq:CH_funct_derivative_S},
\end{eqnarray}
\end{subequations}
where we omitted the position dependence for simplicity.

The Poisson operator ${\mathbf L}^{\rm (1)}$
determines the reversible contributions to the full evolution
equations of the variables ${\mathbf x}$, and hence models the
convective behavior. For our particular choice of variables and in
view of the functional derivatives \eqref{eq:CH_funct_derivative} 
one arrives at the following expression for the Poisson operator
\begin{subequations}\label{eq:CH-L_matrix}
\begin{eqnarray}
 {\mathbf L}^{\rm (1\,)}\left(\mathbf{r},\mathbf{r'}\right)=\left(
\begin{array}{cccc}
    0 & \rho\left(\vrr\right)\frac{\partial \delta}{\partial\vrr} & 0 & 0 \\
    - \frac{\partial \delta}{\partial\vecr}\, \rho\left(\vecr\right) & L^{\rm (1\,)}_{gg} & 
    L^{\rm(1\,)}_{g\epsilon} & 
    -\left(\nabla c\left(\vecr\right)\right)\delta  \\
    0  & L^{\rm (1\,)}_{\epsilon g}& 0 & 0 \\
    0 & -\left(\nabla c\left(\vecr\right)\right) \delta & 0& 0
\end{array}
\right),
\end{eqnarray}
with
\begin{eqnarray}
  L^{\rm (1\,)}_{gg}\left(\vecr,\vrr\right) &=&
  {\mathbf g}\left(\vrr\right)\frac{\partial \delta}{\partial\vrr} 
  -\frac{\partial\delta}{\partial\vecr} {\mathbf g}\left(\vecr\right) ,\label{eq:CH-L_matrix-gg}
\\
  L^{\rm(1\,)}_{g\epsilon}\left(\vecr,\vrr\right) &=&
  -\frac{\partial\delta}{\partial\vecr}\, \epsilon\left(\vecr\right)
  -\frac{\partial \delta}{\partial\mathbf{r}} p\left(\vrr\right)
  , \label{eq:CH-L_matrix-ge} 
\\
  L^{\rm (1\,)}_{\epsilon g}\left(\vecr,\vrr\right) &=&
   \epsilon\left(\vrr\right)\frac{\partial\delta}{\partial\vrr}
  +p\left(\vecr\right)\frac{\partial \delta}{\partial\vrr}  \,, \label{eq:CH-L_matrix-eg} 
\end{eqnarray}
\end{subequations}
where $\delta=\delta\left(\mathbf{r}-\mathbf{r'}\right)$ is Dirac's
$\delta$-function. In addition to the usual hydrodynamic part of the 
Poisson matrix for the variables $\rho$, $\mathbf{g}$, $\epsilon$,
element $L^{\rm (1\,)}_{cg}$ dictates the convection of the configurational 
variable $c$, which is a scalar, so the entry in the Poisson operator 
is as expected by \cite{hco}. The antisymmetry of the Poisson matrix 
\eqref{eq:CH-L_matrix} and the degeneracy requirement are satisfied 
by construction, and the Jacobi identity holds. At this point we mention 
that, in general, the symbol ``$\cdot$'' in \eqref{eq:generic} implies not 
only summation over discrete indices. If field variables are involved 
the operators $\mathbf L$ and $\mathbf M$ are written in terms of 
two space arguments $\left({\vecr},{\vrr}\right)$, and an
integration over ${\vrr}$ must be performed when multiplied
with a function of $\vrr$ from the right. However, in the
case of the field equations being local, one can express $\mathbf L$
and $\mathbf M$ in terms of a single variable $\vecr$ only
\cite{hco}, and no integration is implied when these operators are
multiplied from the right. Such single variable notation is used below for
the friction matrix ${\mathbf M}^{\rm (1\,)}$. 

The irreversible effects in the Cahn-Hilliard
model for a binary fluid under flow arise due to viscosity and heat
conduction, as well as diffusion. Thus, to incorporate them through
the friction matrix ${\mathbf M}^{\rm (1\,)}$, one can write it as
the sum of two matrices $\mathbf{M}^{\rm (1\,)}=\mathbf{M}^{\rm
(1\,),hyd}+\mathbf{M}^{\rm (1\,),dif}$, where $\mathbf{M}^{\rm
(1\,),hyd}$ is the usual hydrodynamic friction matrix, and
$\mathbf{M}^{\rm (1\,),dif}$ contains all the transport coefficients
related to diffusion of mass \cite{EspTh_03,hco}.
The hydrodynamic part of the friction matrix is given in the
$ \vecr $-notation as
\begin{subequations}\label{eq:CH-M-hydrodynamics}
\begin{eqnarray}
\mathbf{M}^{\rm (1\,),hyd}\left(\vecr\right)=\left(
\begin{array}{cccc}
  0 & 0 & 0 & 0 \\
  0 & M^{\rm (1\,),hyd}_{gg}  & M^{\rm (1\,),hyd}_{g\epsilon}  & 0 \\
  0  & M^{\rm (1\,),hyd}_{\epsilon g} & M^{\rm (1\,),hyd}_{\epsilon \epsilon} & 0\\
  0 & 0 & 0 & 0
\end{array}
\right),
\end{eqnarray}
with
\begin{eqnarray}
  M^{\rm (1\,),hyd}_{gg}\left(\vecr \right) &=& 
  -\left(\nabla \eta T \nabla + \mathbf{1}\nabla \cdot \eta T \nabla\right)^{\rm T}
  - \nabla \hat{\kappa}T\nabla, \label{eq:CH-M-hydrodynamics-gg}
\\
  M^{\rm (1\,),hyd}_{g\epsilon}\left(\vecr\right) &=&
  \nabla \cdot \eta T \dot{\bm \gamma}+ 
  \nabla \frac{\hat{\kappa}T}{2}\rm{tr} \dot{\bm{\gamma}}, \label{eq:CH-M-hydrodynamics-ge}
\\
  M^{\rm (1\,),hyd}_{\epsilon g}\left(\vecr \right) &=&
  -\eta T \dot{\bm \gamma}\cdot \nabla
  -\frac{\hat{\kappa}T}{2}\rm{tr}\dot{\bm \gamma}\nabla ,\label{eq:CH-M-hydrodynamics-eg}
\\
  M^{\rm (1\,),hyd}_{\epsilon \epsilon}\left(\vecr \right) &=&
  \frac{\eta T}{2}\dot{\bm \gamma}: \dot{\bm \gamma}+
  \frac{\hat{\kappa}T}{4}\left(\rm{tr}\dot{\bm \gamma}\right)^2-
  \nabla \cdot \lambda^{\rm q}T^2\nabla ,\label{eq:CH-M-hydrodynamics-ee}
\end{eqnarray}
\end{subequations}
with all derivative operators acting on everything to the
right, i.e.,\ also on functions multiplied to the right side of the
operator $\mathbf M$, and with 
with the transport coefficient $\hat{\kappa}$, being a combination
of the dilatational viscosity $\kappa$ and the viscosity $\eta$,
$\hat{\kappa}=\kappa-\frac{2}{3}\eta$, and $\lambda^{\rm q}$ the
thermal conductivity \cite{hco}.

If $M\left(\vecr \right)$ is a position dependent mobility
coefficient, the diffusion part of the friction matrix in the
$ \vecr $-notation is
\begin{subequations}\label{eq:CH-M-diffusion}
\begin{eqnarray}
\mathbf{M}^{\rm (1\,),dif}\left(\vecr \right)=\left(
\begin{array}{cccc}
  0 & 0 & 0 & 0 \\0 & 0 & 0 & 0 \\
  0 &  0 &\mathbf{M}^{\rm (1\,),dif}_{\epsilon\epsilon}
  &  \mathbf{M}^{\rm (1\,),dif}_{\epsilon c}\\
  0 & 0& \mathbf{M}^{\rm (1\,),dif}_{c\epsilon} & \mathbf{M}^{\rm
  (1\,),dif}_{cc}
\end{array}
\right),
\end{eqnarray}
where the $cc$-element is describing the diffusion of $c$, so that
\begin{eqnarray}
 \mathbf{M}^{\rm (1\,),dif}_{cc}\left(\vecr \right)= 
 -\nabla \cdot MT\nabla,\label{eq:CH-M-diffusion-cc}
\end{eqnarray}
and the other three elements are obtained from the symmetry and the
degeneracy conditions
\begin{eqnarray}
  \mathbf{M}^{\rm (1\,),dif}_{c\epsilon}\left(\vecr\right) &=&
  -\nabla \cdot MT\nabla \ke\left(\nabla^2 c\right), \label{eq:CH-M-diffusion-ce}
\\
  \mathbf{M}^{\rm (1\,),dif}_{\epsilon c}\left(\vecr\right) &=&
  -\ke\left(\nabla^2 c\right)\nabla \cdot MT \nabla, \label{eq:CH-M-diffusion-ec}
\\
  \mathbf{M}^{\rm (1\,),dif}_{\epsilon\epsilon}\left(\vecr\right) &=&
  -\ke\left(\nabla^2 c\right)\nabla \cdot MT \nabla\ke\left(\nabla^2 c\right). \label{eq:CH-M-diffusion-ee}
\end{eqnarray}
\end{subequations}
Here again are all variables functions of $\vecr$, and all
derivative operators $\nabla\equiv\frac{\partial}{\partial\vecr}$ acting on
everything to the right, except when they are in brackets and
bounded to act on $c$. The symmetry and the degeneracy requirement
for the matrix $\mathbf{M}^{\rm (1\,),dif}$ are satisfied by
construction.

Finally, by inserting the above building blocks $E$, $S$, $\mathbf L$ 
and $\mathbf M$ into the GENERIC equation \eqref{eq:generic}, the full 
set of time evolution equations takes the form
\begin{subequations}\label{eq:CH_in GENERIC}
\begin{eqnarray}
  \frac{\partial \rho}{\partial t} &=&
  -\nabla\cdot \left( \rho {\mathbf v}\right)
  \label{eq:rhot_in_GENERIC},
\\
  \frac{\partial {\mathbf g}}{\partial t}&=&
  -\nabla\cdot \left({\mathbf v}{\mathbf g}\right)
  -\nabla \cdot \left(\bm{\Pi}+{\bm{\tau}}\right),
  \label{eq:gt_in_GENERIC}
\\  
  \frac{\partial \epsilon}{\partial t} &=&
  -\nabla\cdot \left( \epsilon {\mathbf v}\right)
  -p \nabla \cdot {\mathbf v}
  -\bm{\tau}:\left(\nabla\mathbf{v}\right)^T 
  -\nabla\cdot\mathbf{j}^{\rm q}
\nonumber \\
 && +\ke\left(\nabla^2c\right)\nabla \cdot\left[MT\nabla
  \left(-\frac{\partial s\left(\rho,\epsilon,c\right)}{\partial c}
  -\frac{\ke}{T}\nabla^2c\right)\right], \label{eq:et_in_GENERIC}
\\
  \frac{\partial c}{\partial t} &=& 
  -{\mathbf v}\cdot \left( \nabla c \right)
  +\nabla\cdot\left[MT\nabla \left(-\frac{\partial s\left(\rho,\epsilon,c\right)}{\partial c}
  -\frac{\ke}{T}\nabla^2c\right)\right] \label{eq:ct_in_GENERIC},
\end{eqnarray}
\end{subequations}
where $\bm{\Pi}=\left(p-\frac{1}{2}\ke \left|\mathbf{\nabla}c\right|^2 \right) 
\mathbf{1} +\ke\left(\nabla c\right)\left(\nabla c\right)$ is the pressure tensor,
$\bm{\tau}=\eta \dot{\bm \gamma}-\hat{\kappa}\left(\nabla \cdot \mathbf{v}\right)
\mathbf{1}$ is the viscous stress tensor, and $\mathbf{j}^{\rm q}=-\lambda^{\rm q}
\nabla T$ represents the conductive flow of internal energy. For an isothermal, incompressible flow, these time evolution equations take the form of the standard 
Cahn-Hilliard model. \asja{Similar set of hydrodynamic equations for phase separating
fluid mixtures has been derived from an underlying microscopic dynamics
\cite{EspTh_03}.}

\section{Coarse graining from Cahn-Hilliard to Doi-Ohta level}

A specific feature of the GENERIC formalism is
that it is applicable at different levels of description associated
with different length and time scales.
Each level of description is described by an appropriate
set of state variables and building blocks. The mathematical
structure of GENERIC offers a possibility to perform coarse graining
by focusing not on the time evolution equations, but on the building
blocks $E$, $S$, $\mathbf L$ and $\mathbf M$, see \cite{Ott-07}. 
The transition between
a more detailed (Cahn-Hilliard) level $1$ to a less detailed
(Doi-Ohta) level $2$, which involve the derivation of the
building blocks from level $1$ to level $2$, can then be performed
by systematic procedures \cite{Ilg-09}. The possibility to perform coarse-graining
by focusing on the basic building blocks guarantees that the
thermodynamic structure of the problem will be preserved. 

A systematic coarse-graining procedure is based on the idea of the
separation of time scales. In this way, it is assumed that there
exists a division of the ``fast" and ``slow" degrees of freedom at
the more microscopic level $1$ according to their relaxation times.
By means of the projection-operator technique, one can then
eliminate these ``fast" degrees of freedom from the time evolution
equations for the ``slow" variables. The latter are then associated
to the macroscopic variables at level $2$. The crucial steps in this 
general procedure are to identify the proper mapping of the variables 
of one level to another, $\Pi(x): x \longrightarrow y$, which average 
in a non-equilibrium ensemble $\rho_y\left(x\right)$ is the new
coarse-grained variable $y$
\begin{equation}\label{eq:mapp_definition}
   y=\Big\langle\Pi\left(x\right)
     \Big\rangle_y = \Dx \Pi\left(x\right).
\end{equation}

\subsection{Mappings and ensemble}

For coarse graining from Cahn-Hilliard level, the relevant set of variables 
at the coarse-grained level $y=\left(\rho, {\mathbf g},\epsilon, Q, {\mathbf q}
\right)$ is motivated by the phenomenological Doi-Ohta model 
\cite{DoiOhta_91,WagOttEdw_99}. 
We assume that the hydrodynamic fields $\rho$, $\mathbf g$, and $\epsilon$ 
are smooth on the more microscopic Cahn-Hilliard length scale. Then the 
mappings $\pr$, $\pg$, and $\pe$, simply pick out the hydrodynamic fields,
while we need to determine the relationship of mappings $\pQ$ and $\pq$ 
to the underlying configuration. 
We do this by following the physical meaning rather than the exact 
definitions \eqref{eq:Qq_definition} of the Doi-Ohta variables. 
This is because, first, the interface orientational distribution function 
$f\left(\mathbf{n}\right)$ is not given at the Cahn-Hilliard level, and,
second, due to the difference in modeling of the interface. While in the 
Doi-Ohta model the interface us assumed to be sharp, in the diffuse-interface 
theories, like Cahn-Hilliard one, the interface is given through continuous 
variations of the composition $c$. Therefore, for determining the average 
interfacial area per unit volume $Q$ and its orientation $\mathbf{q}$ one must use the appropriate
combination of the gradients of the composition $\nabla c(\vecr)$, 
and perform ensemble averaging which incorporates ``smoothing" over a 
certain volume. We introduce a smoothing function $\chi(\vecr-\vrr)$, 
which averages the observable over a certain volume $v(\vecr)$ around 
position $\vecr$, and satisfies the normalization condition 
$\irr \,\chi(\vecr-\vrr)=1$. To obtain the variables which
would correspond to the Doi-Ohta averaged interfacial area and its
orientation, volume $v(\vecr)$ must comprise many droplets for
good statistics. That means that the length scale of the smoothing
volume $v(\vecr)$ -- smoothing length $a$ -- must satisfy
\begin{equation}\label{eq:length-scales-comparison}
 \xi \ll L(t) \ll a \ll \Lambda,
\end{equation}
where $\xi$ is a length of the size of the interfacial width, $L(t)$ is the growing
characteristic domain size, and $\Lambda$ is the size of the system.
The first part of the inequality \eqref{eq:length-scales-comparison}, 
$\xi \ll L(t)$, denotes an important fact that we perform the 
coarse-graining procedure only during the late stages of phase separation, 
i.e.,\ when the well defined domains are formed and the coarsening can be 
described by the growth laws given in Section \ref{sec:Introduction}.

Although one can express the coarse-grained variables $Q$ and $\mathbf q$
using gradient of the composition $\nabla c(\vecr)$ and 
average the expression as discussed above, there are still some questions 
which are not solved. As already mentioned, contrary to the 
phenomenological Doi-Ohta model, in the coarse-grained model, the interface 
not only has a finite width, but it does not behave like a zero-width mathematical 
surface. Rather it deforms under the applied flow and its stretching must 
be captured by the new variables, which is not the case in the phenomenological Doi-Ohta model. 
There are different ways to solve these problems. One way would be
that instead of using the Cahn-Hilliard
Poisson operator ${\mathbf L}^{\rm (1\,)}$,
Eq.\eqref{eq:CH-L_matrix}, one needs to use its modification with
the appropriate constraint which would account for the stretching of
the interface. Another possibility would be to make certain
modifications to the original Cahn-Hilliard model in such a way to
make the interfacial width, as well as interfacial tension, fixed.
However, maybe the most straightforward way to solve the presented
problems is to account for the finite interfacial width and its
stretching by introducing an additional variable into our
coarse-grained Doi-Ohta model. This additional variable would be the
average interfacial width $l$. This means that for the 
configurational variables of the coarse-grained Doi-Ohta model we use
new variables $ \{P,{\mathbf p}, l\}$, which are obtained as ensemble 
averages of the suitable mappings $\Pi_P$, $\bm{\Pi}_{\mathbf p}$ and $\Pi_l$. The original 
Doi-Ohta model and the appropriate convective behavior of its variables 
are obtained by transformation of the configurational variables 
$\{P(\vecr),{\mathbf p}(\vecr),l(\vecr)\}$ to 
$\{Q(\vecr),{\mathbf q}(\vecr),l(\vecr)\}$ as
\begin{equation}\label{eq:Qq-Ppl}
  Q(\vecr) = l(\vecr)P(\vecr), \qquad \mathbf{q}(\vecr) = l(\vecr)\mathbf{p}(\vecr).
\end{equation}

We choose the appropriate mappings for the new variables, having in mind 
that the variables $Q$ and ${\mathbf q}$, obtained from $\{P,{\mathbf p},l\}$ 
as \eqref{eq:Qq-Ppl}, should represent the average amount of interface 
per unit volume and its orientation, both of dimension $m^{-1}$. Then, for 
suitable mappings for the new variables we propose
\begin{subequations}\label{eq:CH-to-DO-mappings}
\begin{eqnarray}
  \Pi_P[c](\vecr)  &=& 
  \irr \left|\nabla c(\vrr)\right|^2 \chi(\vecr-\vrr), 
  \label{eq:P-mapping}
\\
  \bm{\Pi}_{\mathbf{p}}[c](\vecr) &=& 
  \irr \left( \left(\nabla c(\vrr)\right)
  \left(\nabla c(\vrr)\right)- \frac{1}{3}\left|\nabla c(\vrr)\right|^2\mathbf{1}\right)
  \chi(\vecr-\vrr), \label{eq:p-mapping}
\\
  \Pi_l[c](\vecr) &=& \frac{\displaystyle \irr \left|\nabla c(\vrr)\right| \chi(\vecr-\vrr)}
  {\displaystyle \irr \left|\nabla c(\vrr)\right|^2\chi(\vecr-\vrr)}. \label{eq:l-mapping}
\end{eqnarray}
\end{subequations}

The next step in the coarse-graining procedure is the choice of the
nonequilibrium ensemble. The natural choice is for it to be of the 
mixed ensemble due to the choice of the macroscopic variables \cite{hco}. 
Since the hydrodynamic fields are simply mapped from the Cahn-Hilliard to 
the Doi-Ohta level, the appropriate probability distribution function is
of the generalized microcanonical type. For the configurational
variables, on the other hand, we choose a generalized canonical
ensemble with the corresponding Lagrange multipliers $\lambda_Q$,
$\bm{\lambda}_{\mathbf q}$ and $\lambda_l$. The total probability measure
$\rho_y[x]$ then takes the form
\begin{subequations}\label{eq:ensemble-CH-to_DO}
\begin{eqnarray}
  \displaystyle \rho_y[x] &=& 
     \delta\left(\Pi_{\rho}- \rho \right)
     \delta\left(\bm{\Pi}_{\mathbf{g}}- \mathbf{g} \right)
     \delta\left(\Pi_{\epsilon}- \epsilon \right)
     \rho_{(Q,\mathbf{q})}[c], 
    \\
     \rho_{(Q,\mathbf{q})}[c] &=& \displaystyle
     \frac{ \Omega_1[x]}{N[y]} 
     \exp \left(-\ir \bm{\lambda}(\vecr):\bm{\Pi}[c](\vecr)\right),
\end{eqnarray}
where the normalization factor $N[y]$ is
\begin{equation}
  N[y]=\int \mathcal{D}c \, \Omega_1[x] 
       \exp \left(-\ir \bm{\lambda}(\vecr):\bm{\Pi}[c](\vecr) \right),
\end{equation}
\end{subequations}
and $\Omega_1[x]=\exp\left(S^{\rm (1)}(x)/\kb\right)$. The projection 
operators $\pQ$, $\pq$, and $\pil$ are, for simplicity, denoted by 
$\bm{\Pi}$ in the above equations. The Lagrange multipliers 
$\lambda_Q$, $\bm{\lambda}_{\mathbf q}$, and $\lambda_l$, denoted by
$\bm{\lambda}$, are determined by the values of the slow variables
$y=\langle x\rangle$ where the average is performed according to
\eqref{eq:mapp_definition}. Their interpretation and the
identification of the proper values for the situation of interest is
difficult and requires dynamic material information \cite{Ilg-09,Mavr_02}. However, 
for the results presented in this paper, the exact form of the nonequilibrium 
ensemble will not be needed.

\subsection{Energy}
The energy of the coarse-grained Doi-Ohta level is obtained by averaging 
the energy of the more detailed Cahn-Hilliard level,
\begin{eqnarray}
    E^{\rm (2)}\left( y\right)= 
         \ir \Dx \irr \chi(\vecr-\vrr)
         \left(\frac{1}{2} 
         \frac{{\mathbf g}\left(\vrr\right)^2}{\rho\left(\vrr\right)}
         +\epsilon \left(\vrr \right)
         +\frac{1}{2}\ke \left|\nabla c(\vrr)
         \right|^2\right).
\end{eqnarray}
Since we assume that the hydrodynamic fields $\rho$,
$\mathbf g$, and $\epsilon$ are smooth on the more microscopic
Cahn-Hilliard length scale, the energy
expression takes the form
\begin{eqnarray}\label{eq:ch-to-do-Energy}
    E^{\rm (2)}\left( y\right)= \ir \left( \frac{1}{2}
    \frac{{\mathbf g}\left(\vecr \right)^2}{\rho\left(\vecr \right)}
    +\epsilon \left(\vecr \right)\right)
    + \ir \frac{\ke}{2l(\vecr)} Q(\vecr),
\end{eqnarray}
where in the last term on the right hand side of the equation, we
have recognized the new variables $Q(\vecr)$ and $l(\vecr)$, determined 
by the mapping \eqref{eq:P-mapping} and \eqref{eq:l-mapping}. The last
term on the right hand side of the above equation describes the energy 
density due to the presence of the interface which is proportional
to the interfacial tension $\sigma$. We therefore obtain the
following well-known expression for the interfacial tension
\begin{equation}\label{eq:CH-to-DO-interfacial tension}
 \sigma(\vecr)=\frac{\ke}{2l(\vecr)},
\end{equation}
which here follows directly as the first result of the performed coarse-graining procedure.

\subsection{Entropy}
The coarse-grained entropy for the case of the 
generalized mixed ensemble $\rho_y[x]$ is obtained from
\begin{equation}\label{eq:DO-entropy}
  S^{\rm (2)}(y) = 
   \Dc \left[
    S^{\rm (1)}(x)
    -\kb \ln \rho_{(Q,\mathbf{q})}[c]
    \right],
\end{equation}
where $S^{\rm (1)}(x)$ is given by \eqref{eq:CH_entropy_functional}.
The additional entropy in the coarse-grained expression, beside
a simple ensemble average of the entropy from the more detailed
level $1$, is associated with the passage from the more microscopic
configurational variable $c$, to the more macroscopic variables $Q$,
$\mathbf q$, and $l$, while the hydrodynamic variables are taken to
the coarser level without affecting the entropy. The additional
entropy takes into account all the microstates with the composition
$c(\vecr)$ consistent with the more coarse-grained state given with
$Q(\vecr)$, ${\mathbf q}(\vecr)$, and $l(\vecr)$. 
\asja{The functional
derivative of the coarse-grained entropy is}
\begin{equation}\label{eq:Sdo_funct_deriv}
  \frac{\delta S^{\rm (2)}(y)}{\delta y}= \left(
 \frac{\partial s\left(\rho,\epsilon, c\right)}{\partial \rho} , {\mathbf 0}\, ,
  \frac{1}{T (\vecr)} \,, \lambda_Q(\vecr),
  \bm{\lambda}_{\mathbf q}(\vecr),\lambda_l(\vecr)
  \right)^{\rm T}.
\end{equation}
\asja{As discussed above, there exists a systematic procedure to obtain the Lagrange 
multipliers from the thermodynamically guided simulations \cite{Ilg-09}. However, their determination is a difficult task and for the calculation of the time friction matrix elements
in section \ref{subsec:irreversible_dynamics} we will assume 
$\bm{\lambda}=\bf{0}$.}

\subsection{Poisson operator}\label{subsec:Poisson_operator}

The general expression for the coarse-grained Poisson operator is 
given by 
\begin{equation}\label{eq:cg_L_operator}
    L^{\rm (2)}\left( y\right)=\left\langle \left(\frac{\delta {\Pi}
    \left(x\right)}{\delta x}\right)^{\rm T}\cdot L^{\rm (1)}\left( x\right)\cdot
    \left( \frac{\delta {\Pi} \left(x\right)}{\delta x}\right) \right\rangle_y.
\end{equation}
which for coarse graining from the Cahn-Hilliard to the Doi-Ohta 
level takes the form
\begin{eqnarray}\label{eq:ch-to-do-L}
  L^{\rm (2)}_{ij}(\vrn{1},\vrn{2}) &=& 
 \Dx \irrn{1}\irrn{2}\,\chi(\vrn{1}-\vrrn{1})\,\chi(\vrn{2}-\vrrn{2}) 
 \nonumber \\
 &&
 \times\irn{3}\irn{4}\frac{\delta \Pi_i[x](\vrrn{1})}{\delta x_k(\vrn{3})}
 L^{\rm (1)}_{kl}(\vrn{3},\vrn{4}) \frac{\delta \Pi_j[x](\vrrn{2})}{\delta x_l(\vrn{4})},
\end{eqnarray}
where the mappings $\bm{\Pi}$ are given by
\eqref{eq:CH-to-DO-mappings}, and the elements of the Poisson
operator $L^{\rm (1)}$ by \eqref{eq:CH-L_matrix}.
To understand the
above expression, note that the integrations over $\vrn{3}$ and
$\vrn{4}$ come from the contraction of the operator $L^{\rm (1)}$
with $\delta\bm{\Pi}/\delta\mathbf{x}$, which includes both matrix
multiplication and integration over the position label. The obtained
quantity must then be averaged in order to obtain the expression at
the Doi-Ohta level. Therefore, the integrations over $\vrrn{1}$ and
$\vrrn{2}$ come from the spatial smoothing of this quantity, since
the ensemble averaging also implies smoothing in space, as noted in
the previous subsection.

From the elements of $L^{\rm (1)}$ in \eqref{eq:CH-L_matrix} and
the mappings $\bm{\Pi}$ in \eqref{eq:CH-to-DO-mappings}, we
conclude that the Poisson operator $L^{\rm (2)}$ has the following
form
\begin{equation}
{\mathbf L}^{\rm (2)}(\vrn{1},\vrn{2})=\left(
\begin{array}{cccccc}
 0 & L^{\rm (2)}_{\rho \mathbf{g}}& 0 & 0 & 0 & 0\\[0.2cm]
 L^{\rm (2)}_{\mathbf{g}\rho} & L^{\rm (2)}_{\mathbf{g}\mathbf{g}}
 & L^{\rm (2)}_{\mathbf{g}\epsilon}& L^{\rm (2)}_{\mathbf{g}P }
 & L^{\rm (2)}_{ \mathbf{g} \mathbf{p}} & L^{\rm (2)}_{\mathbf{g}l }\\[0.2cm]
 0  & L^{\rm (2)}_{\epsilon g}  & 0 & 0 & 0 & 0\\[0.2cm]
 0 &L^{\rm (2)}_{P \mathbf{g}} & 0& 0 & 0 & 0\\[0.2cm]
 0 &L^{\rm (2)}_{\mathbf{p} \mathbf{g}} & 0& 0 & 0 & 0\\[0.2cm]
 0 &L^{\rm (2)}_{l \mathbf{g}} & 0& 0 & 0 & 0\\[0.2cm]
\end{array}
\right).
\end{equation}

In the analytical derivation of the coarse-grained Poisson operator, 
several approximations that are related to the difference in length scales
\eqref{eq:length-scales-comparison} are used. 
The crossover in length scale from the more microscopic Cahn-Hilliard 
level to the more macroscopic Doi-Ohta level is performed through the 
smoothing function $\chi(\vrn{1}-\vrn{2})$. 
While the details of the derivation 
for some of the elements of $L^{\rm (2)}$ are
given in the Appendix A, here we give their final expressions
\begin{subequations}\label{eq:coarse-grained-L}
\begin{eqnarray}
 L^{\rm (2)}_{\rho \mathbf{g}}(\vrn{1},\vrn{2}) &=&
 \rho(\vrn{2})\,\frac{\partial \chi}{\partial \vrn{2}}.
\\
  L^{\rm (2)}_{\mathbf{g}\mathbf{g}}(\vrn{1},\vrn{2}) &=&
  {\mathbf g}\left(\vrn{2}\right)\frac{\partial \chi}{\partial\vrn{2}} 
  -\frac{\partial \chi}{\partial\vrn{1}} {\mathbf g}\left(\vrn{1}\right),
\\
  L^{\rm (2)}_{\epsilon g}(\vrn{1},\vrn{2}) &=&
   \epsilon\left(\vrn{2}\right)\frac{\partial \chi}{\partial\vrn{2}}
  +p\left(\vrn{1}\right)\frac{\partial \chi}{\partial\vrn{2}}  \,,
\\
  L^{\rm (2)}_{P \mathbf{g},\beta}(\vrn{1},\vrn{2}) &=&
 \frac{\partial }{\partial r_{2\alpha}}\left[\left(2p_{\alpha\beta}(\vrn{2})
 -\frac{1}{3}P(\vrn{2})\delta_{\alpha\beta}\right)\chi\right]
 +P(\vrn{2})\frac{\partial \chi}{\partial r_{2\beta}}, \label{eq:L-Pg}
\\
  L^{\rm(2)}_{\mathbf{p} \mathbf{g},\alpha\beta\gamma}(\vrn{1},\vrn{2}) &=&
 \frac{\partial }{\partial r_{2\alpha}}\left[\left(p_{\beta\gamma}(\vrn{2})
 +\frac{1}{3}P(\vrn{2})\delta_{\beta\gamma}\right)\chi\right]
\nonumber \\
  && +\frac{\partial }{\partial r_{2\beta}}\left[\left(p_{\alpha\gamma}(\vrn{2})
  +\frac{1}{3}P(\vrn{2})\delta_{\alpha\gamma}\right)\chi\right]
\nonumber \\
  &&-\frac{2}{3}\delta_{\alpha\beta}\frac{\partial }{\partial r_{2\nu}}
  \left[\left(p_{\nu\gamma}(\vrn{2})+\frac{1}{3}P(\vrn{2})\delta_{\nu\gamma}\right)\chi\right] 
\nonumber \\ 
  && -\left(\frac{\partial }{\partial r_{2\gamma}}p_{\alpha\beta}(\vrn{2})\right)\chi, \label{eq:L-pg}
\\
  L^{\rm(2)}_{l\mathbf{g},\beta}(\vrn{1},\vrn{2}) &=&
 -\frac{l(\vrn{1})}{P(\vrn{1})}p_{\alpha\beta}(\vrn{1})
 \frac{\partial \chi}{\partial r_{2\alpha}}
 -\frac{1}{3}l(\vrn{1})\frac{\partial \chi}{\partial r_{2\beta}}
 -\frac{\partial l(\vrn{1})}{\partial r_{1\beta}}\chi. \label{eq:L-lg}
\end{eqnarray}
\end{subequations}
where $\chi=\chi(\vrn{1}-\vrn{2})$. With this, we obtained all the elements 
of the coarse-grained Poisson operator $L^{\rm(2)}$, since the elements
$L^{\rm(2)}_{\mathbf{g}P}$, $L^{\rm(2)}_{\mathbf{g}\mathbf{p}}$, and
$L^{\rm(2)}_{\mathbf{g}l}$ are obtained from the antisymmetry condition 
for $L^{\rm(2)}$. We note the occurence of the smoothing function 
$\chi(\vrn{1}-\vrn{2})$ in the Poisson operator $L^{\rm(2)}$. Indeed, 
when looking at the elements of $L^{\rm (2)}$ which involve only the hydrodynamic 
fields $\rho$, $\mathbf g$, and $\epsilon$, we see that they differ from the 
appropriate elements of $L^{\rm (1)}$ in \eqref{eq:CH-L_matrix} only in 
locality, i.e.,\ instead of the Dirac delta function, we rather have
a smoothing function $\chi(\vrn{1}-\vrn{2})$. While the original Cahn-Hilliard 
model is local in space, the coarse-grained model, instead, takes into 
account the whole volume element $v(\vrn{1})$ of the smoothing function 
$\chi(\vrn{1}-\vrn{2})$. 
By assumption of the length scales comparison \eqref{eq:length-scales-comparison}, 
this smoothing function behaves simply as a delta function on the Doi-Ohta level 
due to the difference in length scales. 
\asja{Alternatively, the expressions for the elements of the Poisson matrix 
\eqref{eq:L-Pg}-\eqref{eq:L-lg} could be also obtained based only on 
the mathematical character of the variables $P$, $\mathbf{p}$, and $l$. Similar
to the original Doi-Ohta derivation, one would consider the transformation properties 
of the vector $\nabla c$ and the mappings \eqref{eq:CH-to-DO-mappings},
in order to infer the convective behavior.}

The Poisson operator $L^{\rm(2)}$ determines the convective behavior
of the state variables, which can then be compared to the original
Doi-Ohta model. The reversible time evolution equations for the
variables $\{\rho,\mathbf{g},\epsilon,P,\mathbf{p},l\}$ are obtained as
\begin{eqnarray}
 \left.\frac{\partial y_i(\vrn{1})}{\partial t}\right|_{\rm rev}&=&
 \irn{2} L^{\rm(2)}_{ij}(\vrn{1},\vrn{2})
 \frac{\delta E^{\rm(2)}(y)}{\delta y_{j}(\vrn{2})},
\end{eqnarray}
under the previously discussed assumption that the smoothing function 
$\chi(\vrn{1}-\vrn{2})$ is acting as a Dirac delta function 
$\delta(\vrn{1}-\vrn{2})$ at the Doi-Ohta level of description. 
Then, by transformation of the variables, the full set of the 
reversible time evolution equations for the set of state variables 
$\{\rho,\mathbf{g},\epsilon, Q,\mathbf{q},l\}$ takes the form
\begin{subequations}
\begin{eqnarray}
   \left.\frac{\partial \rho}{\partial t}\right|_{\rm rev} &=&
   -\nabla \cdot\left( \rho \mathbf{v}\right),
\\
   \left.\frac{\partial \mathbf{g}}{\partial t}\right|_{\rm rev} &=&
   -\nabla \cdot\left( \mathbf{v} \mathbf{g} \right)
   -\nabla p 
   -\nabla \cdot \Gamma \left(\mathbf{q}-\frac{2}{3}Q\mathbf{1} \right)
   -Q\nabla\Gamma,
\\
   \left.\frac{\partial \epsilon}{\partial t}\right|_{\rm rev} &=&
   -\nabla \cdot\left( \epsilon \mathbf{v}\right)
   -p(\nabla\cdot \mathbf{v}),
\\
   \left.\frac{\partial Q}{\partial t}\right|_{\rm rev} &=&
   -\nabla \cdot \left(Q \mathbf{v}\right)
   -\left(\nabla\mathbf{v}\right)^{\rm T}:\mathbf{q}
   +\frac{2}{3}Q\left(\nabla\cdot\mathbf{v}\right), 
\\
   \left.\frac{\partial \mathbf{q}}{\partial t}\right|_{\rm rev} &=&
   -\left(\nabla\mathbf{v}\right)\cdot \mathbf{q}
   -\mathbf{q}\cdot \left(\nabla\mathbf{v}\right)^{\rm T}
   -\frac{1}{3}Q\dot{\bm{\gamma}}
   -\mathbf{v}\cdot\nabla \mathbf{q} 
   +\frac{1}{3}\mathbf{1}\left(\nabla\mathbf{v}\right):\mathbf{q}
\nonumber \\
   && 
   +\frac{1}{3}Q\mathbf{1}(\nabla\cdot\mathbf{v})   
   +\frac{1}{Q}\left(\mathbf{q}+\frac{1}{3}Q\mathbf{1}\right)
   \left(\mathbf{q}+\frac{1}{3}Q\mathbf{1}\right):(\nabla\mathbf{v}), 
\\
   \left.\frac{\partial l}{\partial t}\right|_{\rm rev} &=&
   -\nabla \cdot \left(l \mathbf{v}\right)
   +\frac{l}{Q}\mathbf{q}:\left(\nabla \mathbf{v}\right)^{\rm T}
   +\frac{4}{3}l\left(\nabla \cdot \mathbf{v}\right). \label{eq:l-convection}
\end{eqnarray}
\end{subequations}
These equations correspond to the convective part of the time
evolution equations of the Doi-Ohta model expressed in the GENERIC
formalism \cite{WagOttEdw_99}. Indeed, when the closure approximation 
\eqref{eq:closure_DO} for the fourth moment of the interfacial normal 
vector $\overline{\mathbf{nnnn}}$ is used, one obtains the exact 
equations as above \asja{(see Eqs. \eqref{eq:Q_DO_convection} and \eqref{eq:q_DO_convection_closure}).} Furthermore, since the Jacobi identity of the 
starting $L^{\rm(1)}$ operator is fulfilled
and the closure approximation we used corresponds to the one made by
Doi and Ohta \cite{DoiOhta_91}, and examined in \cite{WagOttEdw_99},
we assume that the Jacobi identity of the derived Poisson operator
$L^{\rm(2)}$ is valid. However, this assumption might be questioned
due to the approximations used in its derivation and which take
into account different length scales (see Appendix A) .

\subsection{Irreversible dynamics}\label{subsec:irreversible_dynamics}

In this section we turn to the dynamic material properties which are 
contained in a friction matrix. There are two contributions to the 
coarse-grained friction operator, $M^{\rm (2)}=M^{\rm (2)'}+M^{\rm (2)''}$. 
The first contribution, $M^{\rm (2)'}$, is obtained directly by averaging 
the elements of the friction matrix $M^{\rm (1)}$, in the same way as 
for the Poisson operator in \eqref{eq:cg_L_operator}. This direct contribution 
describes dissipative effects that are carried on from a more microscopic 
level of description. 
A second contribution to the coarse-grained friction operator, $M^{\rm (2)''}$, 
results from the processes that are slower than the characteristic time scale 
of the Cahn-Hilliard level (given by the diffusion time $t_{\rm D}$) but are fast
compared to the processes at the Doi-Ohta level of description (with the time 
scale of the interface relaxation time $\tau_{\rm do, 1, 2}$). This contribution 
presents an important feature of the coarse-graining procedure, since it 
captures the additional dissipation arising from the additional processes 
which can be treated as fluctuations on the time scale of the Doi-Ohta level. It 
can be evaluated from the Green-Kubo formula
\begin{equation}\label{eq:Green-Kubo}
     M^{\rm (2)''}_{jk} \left( y\right) =\frac{1}{k_{\rm B}}\int_0^{\tau_{\rm s}} dt
    \Big\langle \dot{\Pi}^{\rm f}_j(x(t)) \dot{\Pi}^{\rm f}_k(x(0))\Big \rangle_y,
\end{equation}
where $\tau_{\rm s}$ separates the times scales between the fast and slow variables,
$\dot{\Pi}^{\rm f}$ is the rapidly fluctuating part of the time derivative of the 
microscopic expressions for the slow variables $y$, and the average is over the 
atomistic trajectories consistent with the coarse-grained state $y$ at $t=0$ and 
evolved according to the microscopic dynamics to the time $t$ \cite{hco}.

In order to identify the main dissipative contribution at the coarse-grained level
and its origin, 
one has to understand all the processes occurring in a phase separation of a 
binary fluid under shear flow. In particular, through numerical simulations we 
tried to understand if there exist any processes in this system that can be 
considered as fast compared to the time scale of the Doi-Ohta level, and to
estimate their contribution to dissipation at the Doi-Ohta level through the 
Green-Kubo part of the friction matrix  $M^{\rm (2)''}$, Eq.\ \eqref{eq:Green-Kubo}.
We performed simulations of the Cahn-Hilliard equation \eqref{eq:ct_in_GENERIC} 
for a binary mixture of volume fraction $\phi=0.3$ 
which was subjected to a shear flow at time $t_0=200t_{\rm D}$ after the 
start of a phase separating process (when the well-defined droplets were formed). 
\asja{We have used the $c^4$-structure for the entropy density $s$, as given in 
subsection \ref{subsec:CH_in_GENERIC}. We used a $1$-dimensional equilibrium 
profile between the two coexisting bulk phases, with the composition field
$c(x)=\sqrt{a/b}\tanh(x/2\xi)$, in order to define the length and time scale using
$\xi=\sqrt{\ke/a}$ and $t_{\rm D}=\xi^2/Ma$.}
Hydrodynamic interaction was not included and, therefore, coarsening of domains proceeded governed by the diffusion process, i.e.,\ through the evaporation-condensation mechanism and the growth law $L(t)\sim t^{1/3}$\cite{Bray-94,Sigg_79,Furuk_85}. 
\asja{Once the well defined interfaces were formed, we looked for the fast processes
permanently at work (like deformation of interface shapes through fast coagulation and break-up, etc.), which would give rise to new irreversible dynamics, going beyond the diffusion and hydrodynamic ones.}

\begin{figure}[t]
\vspace{-0.4cm}
\begin{center}
\includegraphics[scale=0.3,keepaspectratio=true]{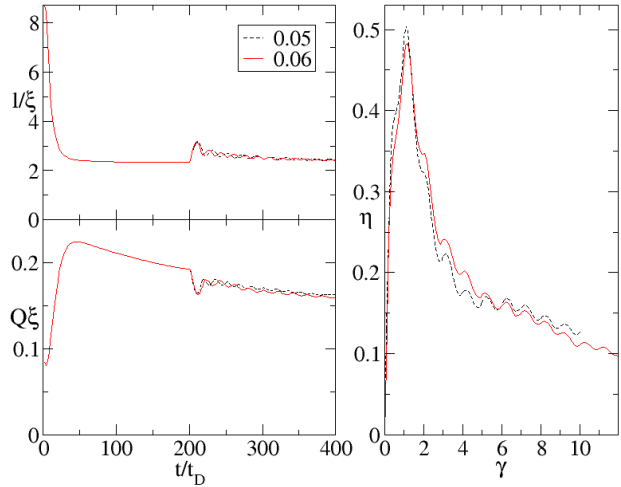}
\end{center}
\vspace{-0.8cm}
 \caption{(Color online) Time evolution of the coarse-grained Doi-Ohta variables for a binary mixture with $\phi=0.3$, subjected to the shear flow at time  $t_0=200t_{\rm D}$ after the beginning of a phase  separating process: interfacial width $l$ (upper left), average amount of interface per unit volume $Q$ (lower left), and shear viscosity, $\eta=-q_{xy}/\dot{\gamma}$ as a function of strain $\gamma=\dot{\gamma}t$ (right). The results for the shear rates $\dot{\gamma}t_{\rm D}=0.05$ and $\dot{\gamma}t_{\rm D}=0.06$ are presented. }
 \label{fig:l and eta for t0=200}
 \vspace{0cm}
\end{figure}

\begin{figure*}[t]
 \vspace{0.0cm}
\begin{center}
\includegraphics[scale=1,keepaspectratio=true]{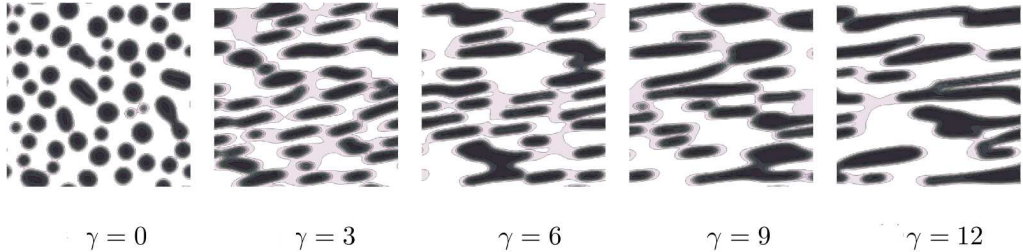}
\end{center}
 \vspace{-0.6cm}
 \caption{Configurations of a binary  fluid with volume fraction  $\phi=0.3$, simulated on
 a $128\times128$ lattice during phase separation under shear flow at different values
 of strain $\gamma=\dot{\gamma}t$. Shear flow was applied after initial time
 of $t_0=200t_{\rm D}$, with the shear rate $\dot{\gamma}t_{\rm D}=0.05$.
 The snapshots were taken until the maximum strain $\gamma=12$. The gray scale 
 is used in order to capture the processes of interface deformation.}
 \label{fig:snapshots-0.3}
 \vspace{0cm}
\end{figure*}

\asja{The time evolution of the coarse-grained Doi-Ohta variables in Fig.\ref{fig:l and eta for t0=200} is computed by a simple use of mappings 
\eqref{eq:CH-to-DO-mappings}.
The shear viscosity, which arises from excess shear stress due to the domain interfaces, is related to the anisotropy element $q_{xy}$ through $\eta=-q_{xy}/\dot{\gamma}$. 
From the time evolution of the average interfacial width $l$, one can see that this 
new structural variable relaxes very quickly and is 
afterwards only affected by the flow through the convective behavior, Eq.\eqref{eq:l-convection}.}
\asja{This numerical analysis shows that the coarse-grained
Doi-Ohta variables fluctuate in time (with the time scale 
of $\sim 1/\dot{\gamma}$)
around their average values due to the competition between the flow
field and the ordering mechanisms, see Fig.\ref{fig:l and eta for t0=200}. 
This can be also seen in the time evolution of the morphology.}
The snapshots of the morphology, presented in Fig.\ref{fig:snapshots-0.3}, 
show that at first  domains are elongated due to the shear flow, which tends to orient
them towards the shear direction. The elongation is then followed by coagulation, 
break-up, shape relaxation, as well as the evaporation-condensation of the droplets,
which are exactly the mechanisms of interfacial relaxation proposed by Doi and 
Ohta \cite{DoiOhta_91}, and Lee and Park \cite{LeePark_94}. Similar interfacial
relaxation processes have been identified in other works which report simulations
of the Cahn-Hilliard model with and without hydrodynamic interaction \cite{OhtaNozDoi_90,CorGonLam-1,CorGonLam-2,CorGonLam-3,CorGonLam-4,Zhang_00,ShouChakr_00,NESS-1,NESS-2,NESS-3,Fielding_08}. 
The morphology
time evolution shows no new processes that are fast compared to the Doi-Ohta level 
of description, and which would emerge as dissipation on the coarse-grained level 
through the Green-Kubo formula \eqref{eq:Green-Kubo}. 
Moreover, the time evolution curves of the Doi-Ohta variables in Fig. 
\ref{fig:l and eta for t0=200} are smooth without the presence of
thermal fluctuations (which were not incorporated into our analysis). 
Since the above described mechanisms 
of interfaces relaxation are also the ones considered for the analysis of different growth 
regimes in the late stage kinetics of phase separation (see Section \ref{sec:Introduction}),
we conclude that the main dissipative contribution at the
coarse-grained level arises through the averaging of the already
present dissipative terms at the Cahn-Hilliard level, $M^{\rm (2)'}$.

\asja{The relaxation of the Doi-Ohta variable $y=\{\rho,\mathbf{g},\epsilon,P,\mathbf{p},l\}$ 
could be expressed from the irreversible part of the GENERIC time evolution equation
\eqref{eq:generic} as}
\begin{eqnarray}\label{eq:DO_irrev_evolution}
 \left.\frac{\partial y_i(\vrn{1})}{\partial t}\right|_{\rm irrev}&=&
 \irn{2} M^{\rm (2)'}_{ij}(\vrn{1},\vrn{2}) \,
 \frac{\delta S^{\rm(2)}(y)}{\delta y_{j}(\vrn{2})}.
\end{eqnarray}
In order to derive the coarse-grained friction matrix  $M^{\rm (2)'}$, we note that the 
Cahn-Hilliard friction matrix $M^{\rm (1)}$ consists of a hydrodynamic and a diffusive 
part, $M^{\rm (1)}=M^{\rm (1),hyd}+M^{\rm (1),dif}$, given in subsection \ref{subsec:CH_in_GENERIC}. 
Therefore  $M^{\rm (2)'}$ will also comprise the hydrodynamic part and a part arising 
from the diffusion. Since the hydrodynamic variables are simply taken over from 
the Cahn-Hilliard level, the appropriate hydrodynamic elements of the new friction 
matrix are the same as in \eqref{eq:CH-M-hydrodynamics}. The diffusive part, 
on the other hand, is related to the configurational variable $c$, and the 
ensemble averaging must be performed in a similar way as for the Poisson operator 
in Section \ref{subsec:Poisson_operator}.
\asja{For the analytical derivation of the coarse-grained  friction matrix $M^{\rm (2)' ,dif }$, in 
analogy to Eq. \eqref{eq:cg_L_operator}, one starts from}
\begin{equation}\label{eq:cg_Mdiff_operator}
    M^{\rm (2)' ,dif }\left( y\right)=\left\langle \left(\frac{\delta {\Pi}
    \left(x\right)}{\delta x}\right)^{\rm T}\cdot M^{\rm (1),dif }\left( x\right)\cdot
    \left( \frac{\delta {\Pi} \left(x\right)}{\delta x}\right) \right\rangle_y,
\end{equation}
\asja{and then employs similar procedure and approximations as in the
derivation of the elements of the Poisson matrix $L^{\rm (2)}$ (see Appendix A).
%
%
For example, the relaxation of the Doi-Ohta variable $Q$ could be expressed 
as an irreversible part of the GENERIC time evolution equation \eqref{eq:DO_irrev_evolution} as}
\begin{equation}
   \left.\frac{\partial Q (\vrn{1})}{\partial t}\right|_{\rm irrev} =
   l(\vrn{1}) \irn{2}M^{\rm (2)',dif }_{P \varepsilon} (\vrn{1},\vrn{2})
   \,\frac{1}{T(\vrn{2})} \label{eq:Q_relax_irr} ,
\end{equation}
\asja{where we have used the approximation
 $\left.\left(\partial Q(\vecr)/\partial t \right)\right|_{irrev} 
\approx l(\vecr) \left.\left(\partial P(\vecr)/ \partial t \right) \right|_{irrev}$,
since the interfacial width $l$ does not change much, compared to the 
other two configurational variables. 
Using the entropy functional derivative \eqref{eq:Sdo_funct_deriv} with 
$\bm{\lambda}=\bf{0}$, results in the matrix element 
$M^{\rm (2)',dif }_{P \varepsilon} $ being the only relevant term 
for the variable $P$ which is left after the averaging \eqref{eq:cg_Mdiff_operator}.
Then, putting the equation \eqref{eq:Q_relax_irr} in the Doi-Ohta equation
\eqref{eq:Q_relax_tau} for the relaxation of $Q$  in a homogeneous case gives 
the following expression for the relaxation time}
\begin{eqnarray}\label{eq:DO_time_relaxation}
\frac{1}{ \tau_{\rm do, 1}}&=& \frac{2\ke M l}{VQ}\,
  \Dc \ir
 \left[ \frac{\partial }{\partial \vecr}
 \left( \frac{\partial^2 c(\vecr)}{\partial \vecr ^2}
 \right) \right]^2,
\end{eqnarray}
where $V$ is the volume of the system. Details of the calculation are given
in Appendix B. Similar
expression could be obtained for the relaxation time $\tau_{\rm do,
2}$. The above formula, gives the expression for the Doi-Ohta
relaxation time $\tau_{\rm do, 1}$ calculated from the
transformation of the element of the friction matrix $M^{\rm (1),dif
}$ containing the relaxation parameters of the Cahn-Hilliard level
at any time $t$. This is the result of the presented systematic
coarse-graining procedure developed within the
GENERIC formalism which shows in which way the Doi-Ohta relaxation
times can be obtained from the more detailed level of description.
However, the above relation is difficult to express analytically in
form of the variables of the Doi-Ohta level, due to the diffusive
terms coming from the Cahn-Hilliard level. On the other hand, the
above expression could, in principle, be calculated numerically
using short time simulations at the Cahn-Hilliard level.
However, the problems concerning the third order derivatives of the
composition field $c$ make this task rather complicated, as well as
the need for determination of the Lagrange multipliers in order to
employ the full GENERIC procedure developed in this paper.

\section{Conclusions and perspectives}

In this paper, we have developed a systematic, thermodynamically guided method which
establishes the coarse-grained Doi-Ohta model, used for rheological
behavior, from the more detailed Cahn-Hilliard model of a phase
separating binary fluid. 
\asja{The main contributions of the present work can be summarized as:
$(1)$ the introduction of the average interfacial width $l$ as an additional 
structural variable; $(2)$ the derivation of the coarse-grained Poisson operator 
$L^{\rm (2)}$, Eq.\eqref{eq:cg_L_operator}, from its finer level analogue; 
$(3)$ no new dissipative processes arise during level jumping
apart from the known hydrodynamic and diffusive dissipation; 
$(4)$ the derivation of an expression \eqref{eq:DO_time_relaxation} 
for the Doi-Ohta relaxation times 
as a function of the elements of the friction matrix describing relaxation 
at the finer Cahn-Hilliard level.}

The crucial step in the coarse-graining procedure was to
identify the relevant coarse-grained variables and find the
appropriate mapping which expresses them in terms of the more
microscopic variables. In order to capture the physics of the
Doi-Ohta level, we introduced the interfacial width $l(\vecr ,t)$ as an additional
variable in the new model. In that way we could account for the
stretching of the interface under flow and derive analytically the
reversible (convective) behavior of the variables $Q(\mathbf{r},t)$
and $\mathbf{q}(\mathbf{r},t)$, which recovers the already
established phenomenological Doi-Ohta model. Introduction of the interfacial 
width $l$ and its time evolution equation, as an addition to $Q$ and 
$\mathbf q$ of the original Doi-Ohta model with zero-width interface, 
offers new possibilities for modeling of the rheological behavior 
of multiphase flows. In addition, the expression 
for the interfacial tension \eqref{eq:CH-to-DO-interfacial tension} 
in terms of the Cahn-Hilliard parameters 
follows as the direct result of the developed systematic coarse-graining 
procedure. 

Considering the irreversible dynamics on the Doi-Ohta level, it has 
been shown that the dissipative processes at this coarse-grained 
level are carried on from the more microscopic Cahn-Hilliard level.
Although their analytical derivation is too complex and rich in structure,
the way to connect the interface relaxation times of the Doi-Ohta
model and the underlying morphology dynamics simulated at the Cahn-Hilliard 
level is established.
Analysis of the numerical investigation of phase separation under 
shear in the diffusive regime revealed no new physical process that 
occur at the time scale which is slow compared to the Cahn-Hilliard 
level and fast from the perspective of the Doi-Ohta level. Therefore, 
there are no new dissipative effects arising in the performed coarse-graining
step through the Green-Kubo type formula, i.e., no new physics
appear. That leads us to the conclusion that all the dissipation has
been introduced into the model by coarse graining from the
reversible atomistic level to the Cahn-Hilliard level. This coarse-graining 
step has been done in the derivation of the hydrodynamic equations of 
a phase separating fluid mixture from the underlying microscopic 
dynamics \cite{EspTh_03}. 
\asja{Furthermore, Espa\'nol and V\'azquez showed in \cite{Esp_02} that all
dissipation is introduced already at the very first coarse-graining step, going from the atomistic to the Fokker-Planck level. From there then follows the conclusion 
that the Cahn-Hilliard level is not so fundamental considering that the
same dissipation mechanisms occur even at the finer levels, and are afterwards 
transmitted to an even more coarse-grained Doi-Ohta level, with no new 
dissipative processes arising.}

The presented results could be used to deal with several interesting open 
problems. First, possible extension of the procedure to more complicated 
phase separating systems with the addition of surfactants or block copolymers
would be very interesting for wider use. Second, the nature of the dissipative 
processes in a phase separating
binary mixture is still far from clear. While during the diffusive regime,
in the absence of thermal noise, we do not recognize any further
fast processes that could be considered as fluctuations at the
Doi-Ohta level, this becomes far from evident during the late stages
of coarsening when inertia becomes important. The nature of the
domain coarsening in the inertial hydrodynamics regime is still
unknown, as well as the asymptotic behavior of a phase separating
system under shear. Furthermore, recent studies showed that a 
nonequilibrium steady state can be reached  at high Reynolds numbers 
(low shear rates) only in the mixed viscous-inertial regime, i.e.,\ with any non-zero 
amount of inertia \cite{NESS-1,NESS-2,NESS-3,Fielding_08}. The fact that no steady state could 
be formed without presence of inertia, no matter how small, suggests that inertia 
plays the role of a singular perturbation in this problem \cite{Fielding_08}.
The formed steady state, arising from a cyclic occurrence of elongation 
of domains followed 
by their break up, coagulation and shape relaxation, is characterized 
by irregular fluctuations around the attained steady state values. It would be 
interesting to understand the nonequilibrium steady state and the 
origin of these fluctuations, using the presented coarse-graining procedure. 
However, the
phenomenological Doi-Ohta theory assumes low Reynolds number, i.e.,
neglects inertia. Deeper insight into the mixed viscous-inertial
regime is therefore crucial in order to develop the connection
between the two levels and to understand the role of inertia in this
problem.

\appendix\section{}

\setcounter{section}{1}
We present a derivation of the elements 
$L^{\rm (2)}_{P \mathbf{g}}(\vrn{1},\vrn{2})$ and
$L^{\rm (2)}_{l \mathbf{g}}(\vrn{1},\vrn{2})$ 
of the coarse-grained Poisson operator $L^{\rm (2)}$ in detail. 
For the element $L^{\rm (2)}_{P\mathbf{g}}(\vrn{1},\vrn{2})$ 
the general expression \eqref{eq:ch-to-do-L} gives
\begin{eqnarray}
  L^{\rm (2)}_{P \mathbf{g}}(\vrn{1},\vrn{2}) &=&
  \Dx \irrn{1}\irrn{2}\chi(\vrn{1}-\vrrn{1})\,\chi(\vrn{2}-\vrrn{2}) 
\nonumber \\
  &&
  \times\irn{3}\irn{4}\frac{\delta \Pi_P[c](\vrrn{1})}{\delta c(\vrn{3})}
   L^{\rm (1)}_{c \mathbf{g}}(\vrn{3},\vrn{4})
  \frac{\delta \bm{\Pi}_{\mathbf{g}}[\mathbf{g}](\vrrn{2})}
  {\delta \mathbf{g}(\vrn{4})},
\end{eqnarray}
which after substitution of the appropriate functional derivatives,
$L^{\rm (1)}_{c \mathbf{g}}$ element, and the integration over
$\vrn{3}$ and $\vrn{4}$ becomes
\begin{eqnarray}
  L^{\rm (2)}_{P \mathbf{g}}(\vrn{1},\vrn{2}) &=&
  \Dc \irrn{1}\irrn{2}\chi(\vrn{1}-\vrrn{1})\,\chi(\vrn{2}-\vrrn{2})
\nonumber \\
  &&
  \times \frac{\partial c}{\partial \vrrn{2}}\;
  \frac{\partial }{\partial \vrrn{2}}\cdot \left( 2\,
  \frac{\partial c}{\partial\vrrn{2}}\, \chi(\vrrn{1}-\vrrn{2})
  \right).
\end{eqnarray}
For further calculations, we use the difference between the Cahn-Hilliard 
and Doi-Ohta length scales \eqref{eq:length-scales-comparison}, so that
$\irrn{1}\chi(\vrn{1}-\vrrn{1})\chi(\vrrn{1}-\vrrn{2})\approx 
\chi(\vrn{1}-\vrrn{2})$ (see \cite{Jelic_09}). Then the last equation, 
after integration over $\vrrn{1}$, becomes
\begin{eqnarray}
  L^{\rm (2)}_{P \mathbf{g}}(\vrn{1},\vrn{2}) = 
 \Dc \irrn{2}\,\chi(\vrn{2}-\vrrn{2})
 \frac{\partial c}{\partial \vrrn{2}}\;
 \frac{\partial }{\partial \vrrn{2}}\cdot \left( 2\,
 \frac{\partial c}{\partial\vrrn{2}}\, \chi(\vrn{1}-\vrrn{2})
 \right).
\end{eqnarray}
After integration by parts, in which surface terms vanish, using
approximation 
$\chi(\vrn{1}-\vrrn{2})\chi(\vrn{2}-\vrrn{2}) \approx
\chi(\vrn{1}-\vrn{2})\chi(\vrn{2}-\vrrn{2})$ valid due to the fact
that the smoothing length is much smaller than the Doi-Ohta length 
scale, equation \eqref{eq:length-scales-comparison}, and the identity
$\partial\chi(\vecr-\vrr)/\partial \vrr=-\partial\chi(\vecr-\vrr)/\partial \vecr$, 
we come to the expression
\begin{eqnarray}
  L^{\rm (2)}_{P \mathbf{g},\beta}(\vrn{1},\vrn{2}) &=&  
  \Dc \times\left\{\frac{\partial }{\partial r_{2\alpha}}\irrn{2}\,2\,
  \frac{\partial c}{\partial r'_{2\alpha}}\,\frac{\partial c}{\partial r'_{2\beta}}\,
  \chi(\vrn{1}-\vrn{2})\chi(\vrn{2}-\vrrn{2})\right.
\nonumber \\
  &&
  -\frac{\partial }{\partial r_{1\beta}}\irrn{2}\,\left|\nabla
  c(\vrrn{2})\right|^2 \chi(\vrn{1}-\vrn{2})\chi(\vrn{2}-\vrrn{2})
\nonumber \\
  &&
  \left.
  -\frac{\partial }{\partial r_{2\beta}}\irrn{2}\,\left|\nabla
  c(\vrrn{2})\right|^2 \chi(\vrn{1}-\vrn{2})\chi(\vrn{2}-\vrrn{2})
  \right\}.
\end{eqnarray}
Finally, with the help of the mappings \eqref{eq:CH-to-DO-mappings}, we can recognize
the appropriate terms in the last equation, so that their ensemble
average gives
\begin{eqnarray}
 L^{\rm (2)}_{P \mathbf{g},\beta}(\vrn{1},\vrn{2})=
 \frac{\partial }{\partial r_{2\alpha}}\left[\left(2p_{\alpha\beta}(\vrn{2})
 -\frac{1}{3}P(\vrn{2})\delta_{\alpha\beta}\right)\chi(\vrn{1}-\vrn{2})\right]
 +P(\vrn{2})\frac{\partial \chi(\vrn{1}-\vrn{2})}{\partial r_{2\beta}}.
\end{eqnarray}

The Poisson operator element
$L^{\rm(2)}_{l\mathbf{g}}(\vrn{1},\vrn{2})$ is obtained under
similar approximations as above, but also the following approximations
\begin{eqnarray}
\irrn{1}\Phi[c](\vrrn{1})\,\chi(\vrn{1}-\vrrn{1})\,\chi(\vrrn{1}-\vrrn{2})
\approx  \Phi[c](\vrn{1})\, \chi(\vrn{1}-\vrrn{2}),
\end{eqnarray}
which holds under the assumption of the difference in length scales
\eqref{eq:length-scales-comparison}, and the assumption
\begin{eqnarray}
 \left\langle\left(\Pi_P[c](\vecr)\right)^n\Pi_l[c](\vecr)\right\rangle \approx 
 \left(\left\langle\Pi_P[c](\vecr)\right\rangle\right)^n
 \left\langle\Pi_l[c](\vecr)\right\rangle,
\end{eqnarray}
for $n=\pm1$. For the element $L^{\rm(2)}_{l\mathbf{g}}(\vrn{1},\vrn{2})$, 
we then obtain
\begin{eqnarray}
  L^{\rm(2)}_{l\mathbf{g},\beta}(\vrn{1},\vrn{2}) &=&
 \frac{1}{P(\vrn{1})}\left\langle\irr\, \frac{\partial c}{\partial r'_{\alpha}}
 \frac{\partial c}{\partial r'_{\beta}} \left|\nabla c(\vrr)\right|^{-1}
 \chi(\vrn{1}-\vrr) \right\rangle
 \frac{\partial \chi(\vrn{1}-\vrn{2})}{\partial r_{2\alpha}}
\nonumber \\
  &&
 -\frac{\partial l(\vrn{1})}{\partial r_{1\beta}}\chi(\vrn{1}-\vrn{2})
 -\frac{2}{3}l(\vrn{1})\frac{\partial \chi(\vrn{1}-\vrn{2})}{\partial r_{2\beta}}
\nonumber \\
  && 
 -2\,\frac{l(\vrn{1})}{P(\vrn{1})}p_{\alpha\beta}(\vrn{1})
 \frac{\partial \chi(\vrn{1}-\vrn{2})}{\partial r_{2\alpha}}.
\end{eqnarray}
In order to express the previous equation in terms of the
coarse-grained Doi-Ohta variables $\{P,\mathbf{p},l\}$, we must use
the following closure approximation for the first term on the right
hand side
\begin{eqnarray}
 && \displaystyle 
 \left\langle\irr\, \frac{\partial c}{\partial r'_{\alpha}}
 \frac{\partial c}{\partial r'_{\beta}} \left|\nabla c(\vrr)\right|^{-1}
 \chi(\vrn{1}-\vrr) \right\rangle \approx 
\nonumber \\
 && \displaystyle 
  C \left\langle\irr\, \frac{\partial c}{\partial r'_{\alpha}}
 \frac{\partial c}{\partial r'_{\beta}} \chi(\vrn{1}-\vrr) \right\rangle
 \left\langle\irr\, \left|\nabla c(\vrr)\right|\chi(\vrn{1}-\vrr)
 \right\rangle \, , 
\end{eqnarray}
where $C$ is chosen in such a way that the previous expression is
valid when its trace is taken. By putting $\alpha=\beta$ into the
above equation, we find
\begin{eqnarray}
 C=\left\langle\irr\, \left|\nabla c(\vrr)\right|^2\chi(\vrn{1}-\vrr)
 \right\rangle ^{-1}.
\end{eqnarray}
Using the above expressions, element
$L^{\rm(2)}_{l\mathbf{g}}(\vrn{1},\vrn{2})$ takes the final form
\begin{eqnarray}
  L^{\rm(2)}_{l\mathbf{g},\beta}(\vrn{1},\vrn{2}) &=&
 -\frac{l(\vrn{1})}{P(\vrn{1})}p_{\alpha\beta}(\vrn{1})
 \frac{\partial \chi(\vrn{1}-\vrn{2})}{\partial r_{2\alpha}}
 -\frac{1}{3}l(\vrn{1})\frac{\partial \chi(\vrn{1}-\vrn{2})}{\partial r_{2\beta}}
\nonumber \\
 && 
  -\frac{\partial l(\vrn{1})}{\partial  r_{1\beta}}\chi(\vrn{1}-\vrn{2}).
\end{eqnarray}

\section{}

We present the derivation of the expression \eqref{eq:DO_time_relaxation}
for the time relaxation $\tau_{\rm do, 1}$. Substitution of the equation \eqref{eq:Q_relax_irr} for the irreversible time evolution of the configurational 
variable $Q$ into the Doi-Ohta relaxation equation 
\eqref{eq:Q_relax_tau} gives 
\begin{eqnarray}\label{eq:DO_time_relaxation1}
\frac{1}{ \tau_{\rm do, 1}}&=& 
 - \frac{l(\vrn{1})}{Q(\vrn{1})} \irn{2}M^{\rm (2)',dif }_{P \varepsilon} (\vrn{1},\vrn{2})
   \,\frac{1}{T(\vrn{2})} \, .
\end{eqnarray}
Under the assumption of a homogeneous system, integration over the whole 
system size of the above expression gives
\begin{eqnarray}\label{eq:DO_time_relaxation_1}
\frac{1}{ \tau_{\rm do, 1}}&=& 
 - \frac{l}{QVT} \irn{1}\irn{2}M^{\rm (2)',dif }_{P \varepsilon} (\vrn{1},\vrn{2}) \, .
\end{eqnarray}
The general expression \eqref{eq:cg_Mdiff_operator} for the coarse-grained 
friction matrix $ M^{\rm (2)',dif }$ gives 
\begin{eqnarray}
 M^{\rm (2)',dif }_{P \varepsilon}(\vrn{1},\vrn{2}) &=&
  \Dx \irrn{1}\irrn{2}\chi(\vrn{1}-\vrrn{1})\,\chi(\vrn{2}-\vrrn{2}) 
\nonumber \\
  &&
  \times\irn{3}\irn{4}\frac{\delta \Pi_P[c](\vrrn{1})}{\delta c(\vrn{3})}
   M^{\rm (1),dif}_{c \epsilon}(\vrn{3},\vrn{4})
  \frac{\delta {\Pi}_{\epsilon}[\epsilon](\vrrn{2})}
  {\delta \epsilon(\vrn{4})},
\end{eqnarray}
which after substitution of the appropriate functional derivatives, $ M^{\rm (1),dif }$
element, and the integration over $\vrn{4}$, $\vrrn{1}$, and $\vrrn{2}$ becomes 
\begin{eqnarray}\label{eq:M2diff_DO}
   M^{\rm (2)',dif }_{P \varepsilon}(\vrn{1},\vrn{2}) &=&
 \Dc \irr \, 2 \ke M T \frac{\partial}{\partial r'_{\alpha}}
 \left(\left(\frac{\partial^2 c}{\partial \vrr^2} \right)
  \chi(\vrn{2}-\vrr)\right)
  \nonumber \\
  &&
  \times
  \frac{\partial^2 }{\partial r'_{\alpha}\partial r'_{\beta}}
  \left(  \frac{\partial c}{\partial r'_{\beta}} \chi(\vrn{1}-\vrr) \right)\, ,
\end{eqnarray}
where we have used the assumption of a homogeneous ($M=const.$)
and isothermal system, as well as the approximations based on a difference
between the Cahn-Hilliard and Doi-Ohta length scales, as in Appendix A.
Substitution of the friction matrix element \eqref{eq:M2diff_DO} into 
equation \eqref{eq:DO_time_relaxation_1},
gives after using the normalization condition $\irr \,\chi(\vecr-\vrr)=1$,
the final formula for the Doi-Ohta time relaxation
\begin{eqnarray}
\frac{1}{ \tau_{\rm do, 1}}&=& \frac{2\ke M l}{VQ}\,
  \Dc \ir
 \left[ \frac{\partial }{\partial \vecr}
 \left( \frac{\partial^2 c(\vecr)}{\partial \vecr ^2}
 \right) \right]^2.
\end{eqnarray}



\begin{thebibliography}{}

\bibitem{Larson_99} R.~G.~Larson, \textit{The {S}tructure and {R}heology of {C}omplex {F}luids} 
(Oxford University Press, New York, 1999).
\bibitem{complex_fluids-1} S.~T.~Hyde and G.~E.~Schroder, {Curr. Opin. Coll. Int. Sci.} {\bf 8}, 5 (2003).
\bibitem{complex_fluids-2} C.~J.~Drummond and C.~Fong, {Curr. Opin. Coll. Int. Sci.} {\bf 4}, 449 (2000).
\bibitem{complex_fluids-3} M.~Caffrey, {Curr. Opin. Sruct. Biol.} {\bf 10}, 486 (2000).

\bibitem{Bray-94} A.~J.~Bray, {Adv. Phys.} {\bf 43}, 357 (1994).
\bibitem{Onuki-97R} A.~Onuki, {J. Phys.: Condens. Matter} {\bf 9}, 6119 (1997).

\bibitem{KrSengHam_92} A.~H.~Krall, J.~V.~Sengers, and K.~Hamano, {Phys. Rev. Lett.} {\bf 69}, 1963 (1992).
\bibitem{Lauger-95} J.~L{\"a}uger, C.~Laubner, and W.~Gronski, {Phys. Rev. Lett.} {\bf 75}, 3576 (1995).
\bibitem{Matsuzaka-98} K.~Matsuzaka, T.~Koga, and T.~Hashimoto, {Phys. Rev. Lett.} {\bf 80}, 5441 (1998).
\bibitem{Derks-08} D.~Derks, D.~G.~A.~Aarts, D.~Bonn, and A.~Imhof, {J. Phys.: Condens. Matter} {\bf 20}, 404208 (2008).
\bibitem{Aarts-05} D.~G.~A.~Aarts, R.~P.~.A.~Dullens, and H.~N.~W.~Lekkerkerker, {New J. Phys.} {\bf 7}, 40 (2005).

\bibitem{OhtaNozDoi_90} T.~Ohta, H.~Nozaki, and M.~Doi, {J. Chem. Phys.} {\bf 93}, 2664 (1990).
\bibitem{PadTox_97} P.~Padilla, and S.~Toxvaerd, {J. Chem. Phys.}  {\bf 106}, 2342 (1997).
\bibitem{CorGonLam-1} F.~Corberi, G.~Gonnella, and A.~Lamura, {Phys. Rev. Lett.} {\bf 81}, 3852 (1998).
\bibitem{CorGonLam-2} F.~Corberi, G.~Gonnella, and A.~Lamura, {Phys. Rev. Lett.} {\bf 83}, 4057 (1999).
\bibitem{CorGonLam-3} F.~Corberi, G.~Gonnella, and A.~Lamura, {Phys. Rev.} E {\bf 61}, 6621 (2000).
\bibitem{CorGonLam-4} F.~Corberi, G.~Gonnella, and A.~Lamura, {Phys. Rev.} E {\bf 62}, 8064 (2000).
\bibitem{Zhang_00} Z.~L.~Zhang, H.~D.~Zhang, and Y.~L.~Yang, {J. Chem. Phys.} {\bf 113}, 8348 (2000).
\bibitem{Berth_01} L.~Berthier, {Phys. Rev.} E {\bf 63}, 051503 (2001).
\bibitem{ShouChakr_00} Z.~Y.~Shou, and A.~Chakrabarti, {Phys. Rev.} E {\bf 61}, R2200 (2000).
\bibitem{NESS-1} A.~J.~Wagner and J.~M.~Yeomans, {Phys. Rev.} E {\bf 59}, 4366 (1999).
\bibitem{NESS-2} P.~Stansell, K.~Stratford, J.~C.~Desplat, R.~Adhikari, and M.~E.~Cates, {Phys. Rev. Lett.} {\bf 96}, 085701 (2006).
\bibitem{NESS-3} K.~Stratford, J.~C.~Desplat, P.~Stansell, and M.~E.~Cates, {Phys. Rev.} E {\bf 76}, 030501(R) (2007).
\bibitem{Fielding_08} S.~M.~Fielding, {Phys. Rev.} E {\bf 77}, 021504 (2008).

\bibitem{Onuki_87} A.~Onuki, {Phys. Rev.} A {\bf 35}, 5149 (1987).
\bibitem{CH-analytical-1} N.~P.~Rapapa and A.~J.~Bray, {Phys. Rev. Lett.} {\bf 83}, 3856 (1999).
\bibitem{CH-analytical-2} N.~P.~Rapapa, {Phys. Rev.} E {\bf 61}, 247 (2000).

\bibitem{DoiOhta_91} M.~Doi and T.~Ohta, {J. Chem. Phys.} {\bf 95}, 1242 (1991).


\bibitem{CaHil_58} J.~W.~Cahn and J.~E.~Hilliard, {J. Chem. Phys.} {\bf 28}, 258 (1958).
\bibitem{model-H} P.~C.~Hohenberg and B.~I.~Halperin, Rev. Mod. Phys. {\bf 49}, 435 (1977).
\bibitem{Sigg_79} E.~D.~Siggia, {Phys. Rev.} A {\bf 20}, 595 (1979).
\bibitem{Furuk_85} H.~Furukawa, {Adv. Phys.} {\bf 34}, 703 (1985).

\bibitem{WagOttEdw_99} N.~J.~Wagner, H.~C.~{\"O}ttinger, and B.~J.~Edwards, {AIChE J.} {\bf 45}, 1169 (1999).
\bibitem{LeePark_94} H.~M.~Lee and O.~O.~Park, {J. Rheol.} {\bf 38}, 1405 (1994).
\bibitem{DO-different-1} M.~Grmela and A.~Ait-Kadi, {J. Non-Newtonian Fluid Mech.} {\bf 77}, 191 (1998).
\bibitem{DO-different-2} M.~Grmela, A.~Ait-Kadi, and L.~A.~Utracki, {J. Non-Newtonian Fluid Mech.} {\bf 77}, 253 (1998).
\bibitem{DO-different-3} C.~Lacroix, M.~Grmela, and P.~J.~Carreau, {J. Rheol.} {\bf 42}, 41 (1998).
\bibitem{DO-different-4} M.~Bousmina, M.~Aouina, B.~Chaudhry, R.~Guenette, and R.~E.~S.~Bretas, {Rheol. Acta} {\bf 40}, 538 (2001).
\bibitem{DO-different-5} I.~Vinckier and H.~M.~Laun, {J. Rheol.} {\bf 45}, 1373 (2001).
\bibitem{DO-different-6} J.~F.~Gu and Miroslav Grmela, {Phys. Rev. E} {\bf 78}, 056302 (2008).

\bibitem{hco} H.~C.~{\" O}ttinger, \textit{Beyond Equilibrium Thermodynamics} (Wiley, Hoboken, N.J., 2005).
\bibitem{OttGrm-97a} M.~Grmela and H.~C.~{\" O}ttinger, {Phys. Rev. E} {\bf 56}, 6620 (1997).
\bibitem{OttGrm-97b} H.~C.~{\"O}ttinger and M.~Grmela, {Phys. Rev. E} {\bf 56}, 6633 (1997). 

\bibitem{Onu_05} A.~Onuki, {Phys. Rev. Lett.} {\bf 94}, 054501 (2005).

\bibitem{Jelic_09} A.~Jeli\'c, {PhD thesis} (ETH Zurich, Switzerland, 2009).

\bibitem{EspTh_03} P.~Espa{\~n}ol and C.~Thieulot, {J. Chem. Phys.} {\bf 118}, 9109 (2003).

\bibitem{Ott-07} H.~C.~{\"O}ttinger, {MRS Bulletin} {\bf 32}, 936 (2007). 

\bibitem{Ilg-09} P.~Ilg, H.~C.~{\"O}ttinger and M.~Kr\"oger, {Phys. Rev. E} {\bf 79}, 011802 (2009). 

\bibitem{Mavr_02} V.~G.~Mavrantzas and H.~C.~{\"O}ttinger, {Macromolecules} {\bf 35}, 960 (2002).

\bibitem{Esp_02} P.~Espa{\~n}ol and F.~V\'azquez, {Phil. Trans. R. Soc. Lond. A } {\bf 360}, 383 (2002).

\end{thebibliography}

\end{document}